\documentclass[]{spie}  

\usepackage{amsmath,amsfonts,amssymb}
\usepackage{graphicx}
\usepackage[colorlinks=true, allcolors=blue]{hyperref}
\usepackage[table]{xcolor}
\usepackage{subcaption}

\def\fermi{{\it Fermi}}

\title{AMEGO: Exploring the Extreme Multimessenger Universe}

\author[a]{Carolyn A. Kierans}
\author[b]{the AMEGO Team}
\affil[a]{NASA Goddard Space Flight Center, 8800 Greenbelt Rd, Greenbelt, Md, USA}
\affil[b]{https://asd.gsfc.nasa.gov/amego/team.html}

\authorinfo{Further author information: \\
E-mail: carolyn.a.kierans@nasa.gov, Telephone: (301) 286-7628\\ }

\begin{document} 
\maketitle
\begin{abstract}
The All-sky Medium Energy Gamma-ray Observatory (AMEGO) is a Probe-class mission concept that will provide essential contributions to multimessenger astrophysics in the next decade. AMEGO operates both as a Compton and pair telescope to achieve unprecedented sensitivity between 200~keV and $>$5~GeV. The instrument consists of four subsystems. A double-sided strip silicon Tracker gives a precise measure of the first Compton scatter interaction and tracks of pair-conversion products. A novel CdZnTe Low Energy Calorimeter with excellent position and energy resolution surrounds the bottom and sides of the Tracker to detect the Compton-scattered photons which enhances the polarization and narrow-line sensitivity. A thick CsI High Energy Calorimeter contains the high-energy Compton and pair events. The instrument is surrounded by a plastic anti-coincidence detector to veto the cosmic-ray background. We have performed detailed simulations to predict the telescope performance and are currently building a prototype instrument. The AMEGO prototype, known as ComPair, will be tested at the High Intensity Gamma-Ray Source in 2021, followed by a balloon flight in Fall of 2022. In this presentation we will give an overview of the science motivation, a description of the observatory, and an update of the prototype instrument development.

\end{abstract}


\section{INTRODUCTION}
\label{sec:science}

The All-sky Medium Energy Gamma-ray Observatory (AMEGO) is a Probe-class mission concept that will provide groundbreaking new capabilities for multimessenger astrophysics.
AMEGO will cover the energy range from 200~keV to $>$~5~GeV with unprecedented sensitivity, enhanced by a wide field of view and capabilities in gamma-ray polarimetry and narrow-line spectroscopy.
By studying sources in our extreme Universe, from compact objects to active galactic nuclei, AMEGO will provide essential electromagnetic observations of multimessenger sources.


Gamma-ray observations have played a critical role in the detection and identification of all multimessenger sources to date. Gamma rays from SN1987A\cite{sn1987a}, GW170817\cite{grb170817}, and TXS 0506+056\cite{txs0506} have not only provided the electromagnetic counterpart for identification, but analysis of the gamma-ray signal has led to unique insights not attainable in other energy bands.
The next generation of gravitational wave (GW) and neutrino detectors will significantly increase the number of source detections with a sensitivity beyond what the current gamma-ray missions can match; there needs to be a similar advancement in gamma-ray missions to provide the essential electromagnetic signature of these sources.

AMEGO will be a key contributor to the next-generation revolution in multimessenger astrophysics by studying the astrophysical objects that produce GW and neutrinos to answer compelling science questions in the extreme Universe. 
The three main science goals of AMEGO are to:
\begin{itemize}
    \item Understand the physical processes in the extreme conditions around compact objects involved in gravitational wave events and other energetic phenomena
    \item Resolve the processes of element formation in extreme environments, such as kilonovae and supernovae
    \item Decipher the operating processes of jets in extreme environments such as gamma-ray bursts and active galactic nuclei
\end{itemize}
In the past few decades, technological developments have significantly improved the performance of recent MeV gamma-ray missions.
Table~\ref{tab:properties} summarizes the expected performance of a few key telescope parameters for AMEGO. These will be further discussed in Sec.~\ref{sec:performance}.

In addition to the groundbreaking multimessenger astrophysics achievable with AMEGO, the mission will provide unprecedented sensitivity in the historically under-explored MeV gamma-ray regime. 
AMEGO is a general-purpose observatory that will provide new discovery capabilities across four orders-of-magnitude in energy, covering the range from X-ray to GeV missions. 
Combining exceptional continuum sensitivity, advanced nuclear line spectroscopy, and gamma-ray polarimetry, AMEGO will contribute unique measurements of the extreme universe and provide excellent synergies with observations at other wavelengths.

\begin{table*}[tb]
\centering
\small
\rowcolors{1}{blue!9!white!97!green}{blue!6!white!98!green}
\caption{AMEGO has been optimized for excellent flux sensitivity, broad energy range, and large field of view to enable the study of multimessenger sources.}
\vspace{2mm}
\begin{tabular}{|l|l|}
\hline
{\bf Energy Range} & 200 keV to $>$5 GeV \\
{\bf Angular Resolution per Photon} & 2.5$^\circ$ (1 MeV), 2$^\circ$ (100 MeV), $1^\circ$ (1 GeV)\\
{\bf Energy Resolution} & $1\%$ (1~MeV, FWHM/E), $\sim$10$\%$ (1 GeV, FWHM/E) \\
{\bf Field of View} & 2.5 sr (20\% of the sky) \\
{\bf Continuum Sensitivity } & $7\times10^{-12}$ (1 MeV), $3.5\times10^{-12}$ (100 MeV) erg cm$^{-2}$ s$^{-1}$ in 5 years\\
{\bf Line Sensitivity} & $1\times 10^{-6}$ ph cm$^{-2}$ s$^{-1}$ for the 1.8 MeV $^{26}$Al line in 5 years\\ 
{\bf Polarization Sensitivity} & 4\% MDP for a 100 mCrab flux, observed for 10$^6$ s \\
\hline
\end{tabular}
\label{tab:properties}
\end{table*}

The AMEGO mission concept has been supported by over 200 international scientists and was submitted to the Astro2020 Decadal Survey~\cite{AMEGOBAAS,AMEGORFI}.
For a more detailed discussion of AMEGO's science capabilities, refer to Refs.~\citenum{AMEGOBAAS}, \citenum{AMEGORFI}, and the list of relevant Astro2020 White Papers (\url{https://asd.gsfc.nasa.gov/amego/science.html}).

In this paper we will give a detailed description of the AMEGO telescope in Sec.~\ref{sec:instrument}. Detailed simulations have been used to predict the performance of AMEGO, and these are presented in Sec.~\ref{sec:performance}. Finally, in Sec.~\ref{sec:prototype}, we introduce the work being done to build a prototype AMEGO instrument and give an update on the development.

\section{THE AMEGO INSTRUMENT}
\label{sec:instrument}

Compton scattering is the dominant photon-matter interaction from a few hundred keV to $\sim$10 MeV (depending on the scattering material), while pair conversion dominates at higher energies. 
AMEGO takes advantage of these two interaction mechanisms to detect gamma rays with unprecedented sensitivity across a wide field of view (FOV) and four orders of magnitude in energy. 
To motivate the AMEGO instrument design, we will present an overview the design considerations for the detection of both relevant interaction types.

The interaction cross-section for photons and matter depends on the atomic number (Z) of the material. 
To enhance the Compton scattering cross-section, and to minimize the effect of Doppler broadening on the angular resolution~\cite{zoglauer2003}, a low-Z detector material should be used for the first Compton scatter interaction. 
However, the photon must be fully absorbed to measure the total energy of the event, so a high-Z material should then be used around the scattering detector to enhance the chance of photoabsorption for the Compton-scattered photon. 
In the pair regime, high-Z materials promote pair-conversion, while thin layers of a low-Z detector material are ideal for tracking the pair-products.
Typically this is handled by combining layers of high-Z conversion foils with position sensitive detectors, similar to the design on the \fermi-LAT~\cite{fermi}.
However, these high-Z foils limit the angular resolution and energy measurements at lower energies due to multiple scattering; therefore, a pair telescope optimized for the MeV range should use the tracker detector material itself as the converter.
The pair event must be fully contained to determine the energy, so a thick high-Z absorber detector should be used under the tracking layers to measure the initial gamma-ray energy. 

In a Compton scattering event, a gamma ray deposits its energy in two or more discreet interaction points that must be measured with high precision to reconstruct the origin of the photon. 
In pair-conversion, the gamma ray converts into an electron-positron pair whose tracks through the detector material can be used to determine the initial gamma-ray direction. Both interactions require fine position resolution (typically $\sim$mm) in the scatterer or tracker, while the resolution elements of the absorber detector are not as critical to the reconstruction.
Additionally, Compton event reconstruction necessitates a precise measure of the energy deposited (typically $\Delta E/E$ of a few \%) in the scatterer, and both event types require a spectrally-sensitive absorber. 

\begin{figure}[b]
    \centering
    \includegraphics[width=3in]{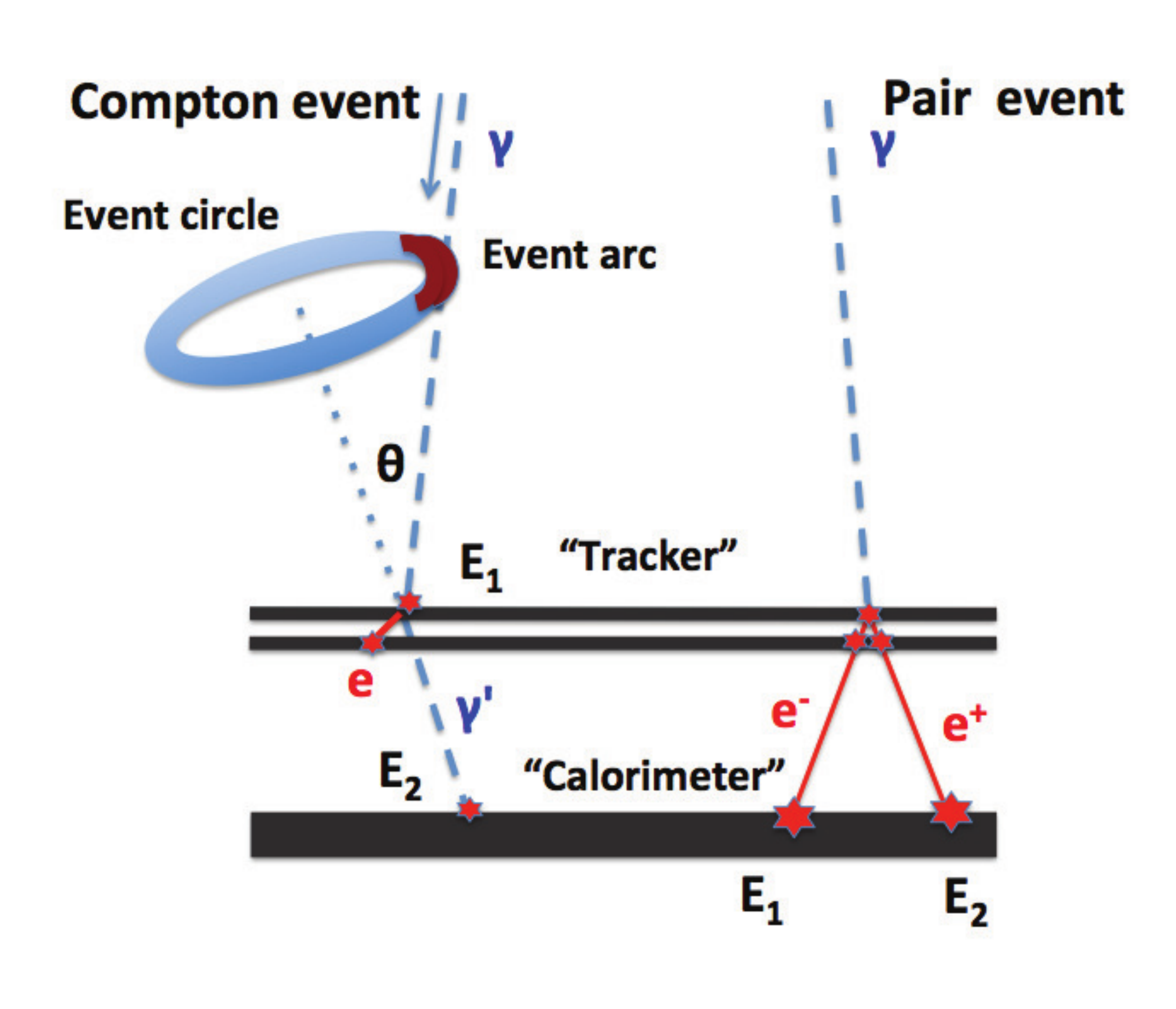}
    \caption{AMEGO detects gamma rays through both Compton scattering and pair production. At energies less than $\sim$10~MeV, a photon will predominantly Compton scatter, and a precise measure of the position and energy of each interaction can constrain the initial photon direction to a circle on the sky. If the direction of the Compton-scattered electron is measured, this circle can be reduced to an arc, which reduces the background contribution. At energies greater than $\sim$10~MeV, a gamma ray will undergo pair conversion, and the track of the electron-positron pairs in the instrument determine the initial photon direction. The energy is measured in the calorimeter.}
    \label{fig:ComptonPair}
\end{figure}

When these design considerations are combined, a hybrid telescope that is capable of detecting both interaction types that consists of a low-Z Tracker, which serves as the scattering element for Compton events and converter for pair events, and high-Z Calorimeter, to measure the energy, is ideal.
Figure~\ref{fig:ComptonPair} illustrates how a Tracker and Calorimeter geometry work to detect both event types. 
At energies less than $\sim$10~MeV, a photon will predominantly Compton scatter in the Tracker layers, and the Compton-scattered photon will then be absorbed in the Calorimeter.
By a precise measure of the interaction position and energy deposited in the Tracker and Calorimeter, the standard Compton equation can be used to determine the Compton scatter angle $\theta$ of the first interaction, which constrains the initial photon direction to a circle on the sky called the ``event circle.''
An added advantage of using a tracking detector is that the direction of the Compton-scattered electron can be measured if it subtends multiple layers, and with this additional kinematic information, the initial direction of the gamma ray can be constrained to an arc on the sky, as opposed to a circle. 
At energies greater than $\sim$10~MeV, a gamma ray will predominantly undergo pair conversion in the Tracker, and the path of the electron-positron pair can be tracked to determine the initial direction of the photon. 
The Calorimeter contains the electromagnetic shower that develops as the electron-positron pair enter the high-Z detector volume to give the total energy in each event. Additionally, the position sensitive absorber allows for the shower profile to be imaged, which provides an important background discriminator and allows for reconstruction of events that are not fully contained.

Observations in the MeV regime are background dominated; therefore, any reduction in background can dramatically increase the sensitivity of the instrument. The first line of defence again the background in orbit is an anti-coincidence detector (ACD) which vetoes interactions in the detector from cosmic-rays. 
At the lower energies, one of the dominant background sources is activation of the instrument from the bombardment of these cosmic rays. 
An additional challenge in the MeV range is the adverse effects of passive material; interactions in passive material render Compton events useless and affect the energy and angular measurements for pair events.
By reducing the amount of passive material near the detector, one can minimize these effects.


\begin{figure}[tb]
    \centering
    \includegraphics[height=2.8in]{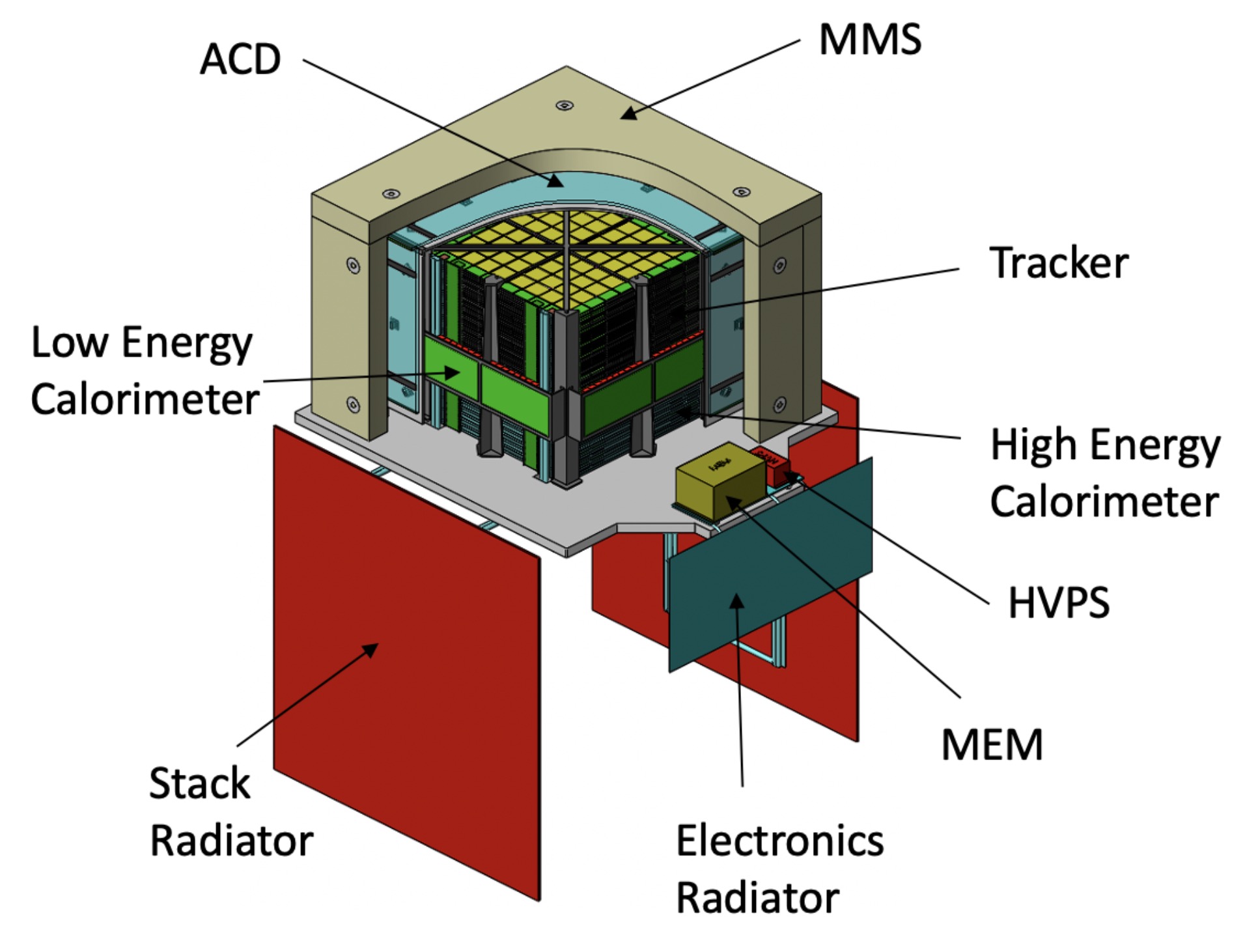}
    \hfill
    \includegraphics[height=2.8in]{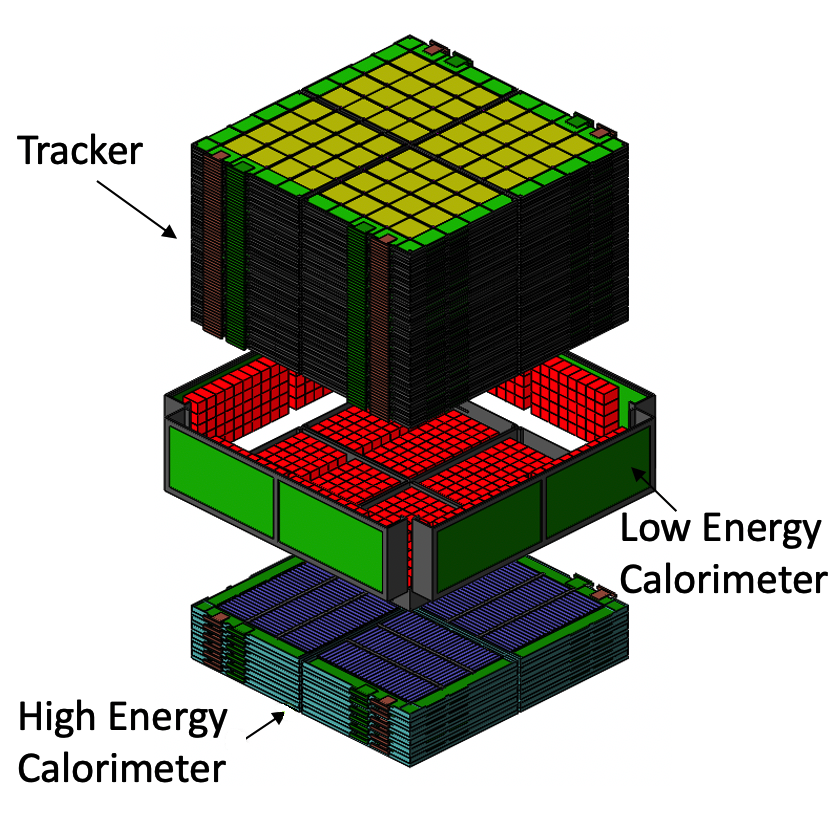}

    \caption{The AMEGO instrument consists of four subsystems: the silicon Tracker, the CZT Low Energy Calorimeter, and the CsI High Energy Calorimeter, which are all surrounded by the ACD. \textit{Left} The ACD and the micrometeoroid shield (MMS) are cut away to expose the tower structure of AMEGO. The full instrument measures 1.6$\times$1.6$\times$1.2~m and includes a Main Electronics Module (MEM), a high voltage power supply (HVPS), and radiators. \textit{Right} An exploded view shows the three inner detector subsystems, where the Low Energy Calorimeter surrounds the bottom third of the Tracker.}
    \label{fig:amego}
\end{figure}

With the above considerations, AMEGO has been designed to consist of four main subsystems, as shown in Fig.~\ref{fig:amego}.
Photons will first interact in the 60 layers of double-sided silicon strip detectors (DSSDs), which form the silicon Tracker, and acts as the scatterer and converter for Compton and pair events, respectively. 
DSSDs are chosen as the Tracker detector element because they provide excellent 3D position resolution, good energy resolution, and have a high technology readiness level (TRL). 
Surrounding the bottom of the Tracker sits the cadmium zinc telluride (CZT) Low Energy Calorimeter, which has excellent spatial and spectral resolution to measure Compton-scattered photons and low-energy pair showers. 
This novel subsystem also operates as a stand-alone Compton detector, not requiring a first interaction in the Tracker, which dramatically increases detection efficiency of AMEGO at the lowest energies.  
At the bottom of the instrument, the cesium iodide (CsI) High Energy Calorimeter measures the electromagnetic shower from high-energy pair events.
Finally, these detector systems are surrounded by a plastic scintillator ACD that rejects the charge-particle background in orbit. 
This Tracker/Calorimeter geometry using silicon and CsI, which is similar to the \fermi-LAT but optimized for lower energies, has been proposed for various MeV telescope concepts over the past two decades\cite{mega, tigre, eastrogam} and is a well-understood design.

For ease of integration and assembly, the AMEGO instrument is composed of four identical towers that make up the four quadrants of the instrument. 
The electronics readout for each detector system is placed at the outer edges of these towers to reduce the amount of passive material within the detector volume. 
Furthermore, each tower has a modular design that consists of identical detector elements to save cost and simplify construction and assembly. 
All primary structural elements that support the detectors are designed using low-Z carbon composite materials to reduce activation of the instrument.
The four AMEGO subsystems and the observatory operations will be described in further detail in the following sections.

\subsection{Tracker}
\label{sec:AMEGOTracker}

The AMEGO Tracker subsystem needs to have good spectral resolution and excellent position resolution to measure the first Compton scatter interaction and the track of pair-conversion products.
Unlike the \fermi-LAT, which uses alternating layers of single-sided silicon strip detectors to track particles, AMEGO needs to determine the position of interaction within a single layer in the Compton regime, and thus requires double-sided strip detectors (DSSDs). 
The specific design choices for the Tracker, such as the separation between layers, thickness of the silicon, and the strip pitch, are determined by the instrument requirements. 
The angular resolution in the pair regime, for example, is driven by the thickness, position resolution, and separation of the DSSDs, while the distance between layers additionally drives the FOV.
Furthermore, passive material in the Tracker can be detrimental to accurate Compton and low-energy pair event reconstruction, and therefore must be kept to an absolute minimum.
These considerations have been taken into account for the AMEGO Tracker design to optimize the continuum sensitivity for high-energy Compton events and low-energy pair events.

\begin{figure}[tb]
\centering
\begin{minipage}{.48\textwidth}
    \centering
    \includegraphics[height=2.3in]{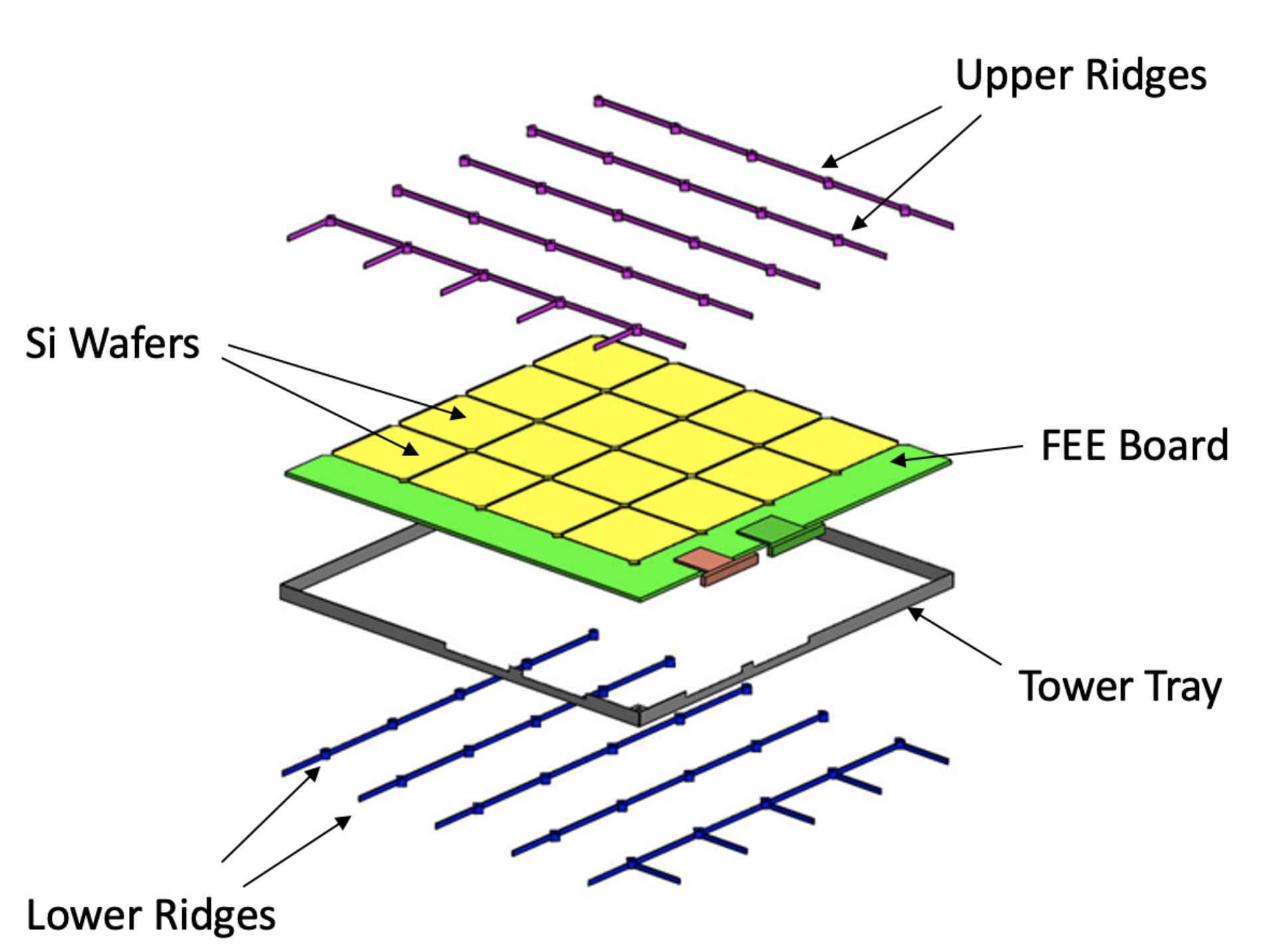}
    \caption{A single layer of the AMEGO Tracker consists of a $4 \times 4$ array of DSSDs, which are daisy-chained and read out through the FEE boards on the outer edges of the array. 
    Mechanically, ridges of low-Z composite supports the edge of each wafer and minimizes the passive material in the detector area. This is only one quadrant of the full Tracker detector.}
    \label{fig:AMEGOTracker}
\end{minipage}%
\hfill
\begin{minipage}{.48\textwidth}
    \centering
    \includegraphics[height=2.3in]{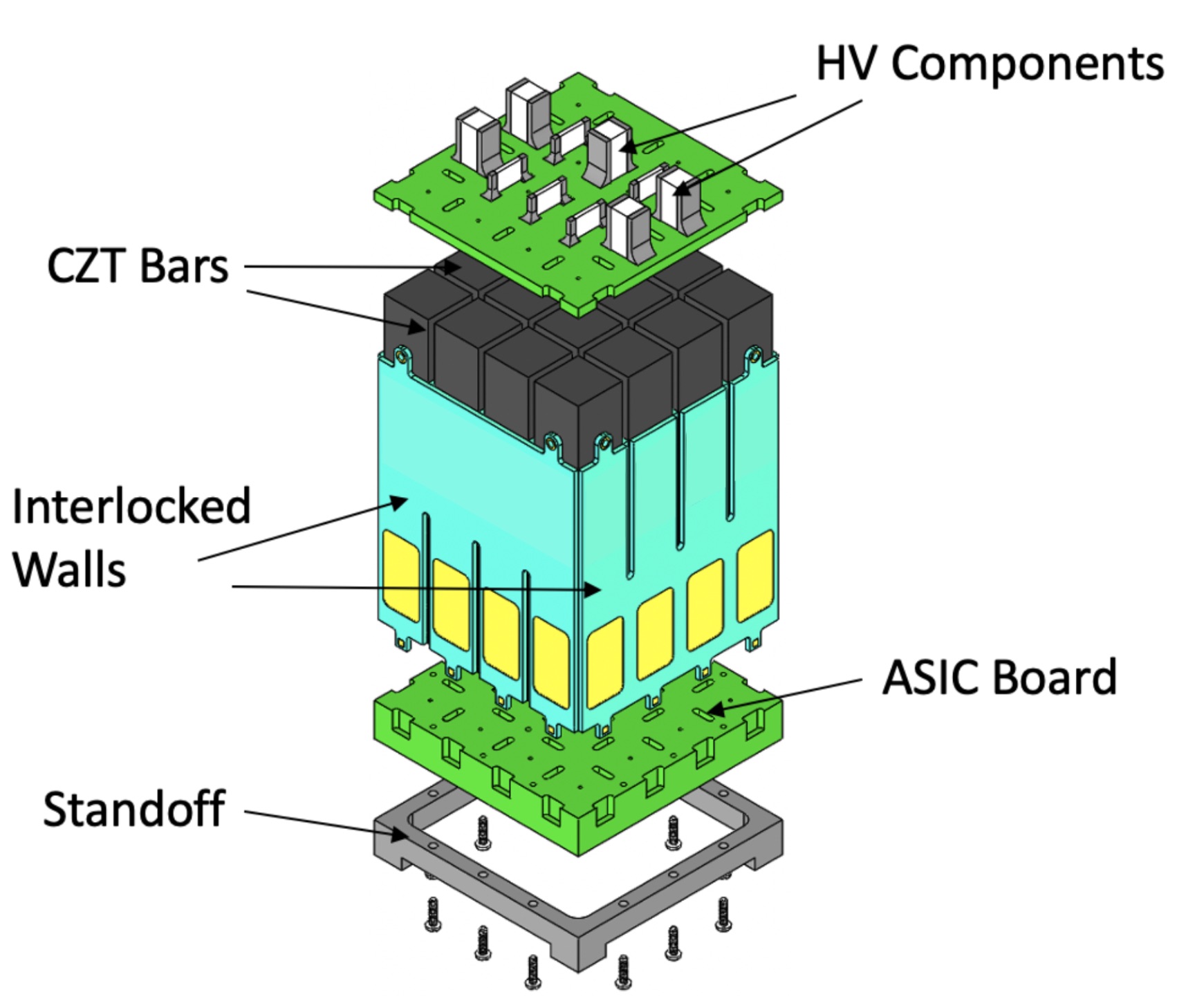}
    \caption{The AMEGO Low Energy Calorimeter consists of a base array of 16 CZT bars, as shown here. Each array is enclosed via a interlocking printed circuit board (PCB) structure. The top detector board routes an external high voltage (HV) to the detector array, while the readout ASIC is mounted to the bottom board.}
    \label{fig:AMEGOCZT}
\end{minipage}
\end{figure}

The AMEGO silicon Tracker consists of 60 layers of 0.5~mm thick DSSDs. 
Each 9.5~cm square wafer has a 500~$\mu$m strip pitch for the 190 orthogonal strips on the front and back side of the detector. 
The wafers are daisy-chained with wire bonds in a $4 \times 4$ array, making a single layer of 16 wafers, as shown in Fig.~\ref{fig:AMEGOTracker}.
This chaining decreases the number of readout channels and brings all of the front-end electronics to the edge of the tower. 
The layers of the Tracker have a 1.0~cm separation. 

The wafers are supported in a grid of carbon fiber reinforced polymer material, as shown in Fig.~\ref{fig:AMEGOTracker}. 
This ridged structure minimizes the passive material in the Tracker volume by supporting only the edge of each wafer where the wirebonds are made. 
Each full layer is further supported in a mounting tray, where a rigid structure is achieved through the use of alignment pins between each layer as opposed to interface screws to simplify integration.
The four towers of the AMEGO instrument are additionally supported by corner posts and a frame structure also made of low-Z composite material. 
Preliminary Finite Element Modeling analysis indicates the support far exceeds the structure requirements imposed by the expected launch loads.

As both the low-energy pair and Compton regime require an accurate measure of the energy deposited in each layer, an analog readout is required. Each daisy-chained strip (4 wafers together) is read out at the FEE board at the edge of the layer by ASICs, where the IDEAS VATA460.3 is currently baselined.
The corner of each layer contains a digital backend board with an FPGA that controls all ASICs in a single layer and communication with the rest of the instrument.
Altogether, the AMEGO instrument has 364,800 Tracker electronics channels, and the heat generated in these electronics is dissipated through heat straps that connect to each layer FEE board. 

The Tracker is based off of heritage from \fermi-LAT and other gamma-ray and cosmic-ray missions. 
DSSDs have been previously flown in AMS-02~\cite{ams02}, Astro-H~\cite{astrohHXI} and Pamela~\cite{pamela}, and, with a similar silicon tower structure, much of the integration and testing will be based off of the \fermi-LAT instrument. We have assessed the AMEGO Tracker to have TRL 6.

\subsection{Low Energy Calorimeter}
\label{sec:AMEGOCZT}

The performance in the Compton regime can be enhanced by a calorimeter with excellent spectral and spatial resolution.
The function of this subsystem is two fold: to precisely measure the Compton-scattered photon from interactions in the silicon, and to increase the effective area for the low energy Compton events which only interact in the calorimeter and have superior spectral information.
This high-precision detector needs to be thick ($\sim$cm) to provide stopping power for MeV photons, and 
the coverage of this calorimeter up the sides of the Tracker is a balance between detecting large Compton scatter angles and maintaining a large FOV.
The AMEGO Low Energy Calorimeter has been optimized for narrow-line sensitivity and effective area in the Compton regime.

The instrument requirements are satisfied with a thick segmented semiconductor detector.
Brookhaven National Laboratory (BNL) has recently developed thick CZT detectors that achieve excellent spatial and spectral resolution \cite{bolotnikov2020, bolotnikov2016}, while operating at room temperature.
By using a large geometric ratio detector and placing a grounded electrode near the anode, the virtual Frisch-grid effect shields the anode from the holes.
Analogous to a drift chamber, the relative anode and cathode signals give the depth of interaction within the thick detector volume. 
The novel aspect of the BNL work has been the segmentation of this virtual Frisch-grid electrode into a four sensing electrode; using the relative amplitude of the induced charge on each side pad, the X-Y position of interaction within the detector can be measured.
This additional position information can in turn be used to make correction for deformities in the CZT crystal structure and improve upon the energy measurement. 
Laboratory measurements have shown $<$1\% energy resolution at 662 keV, and a sub-mm position resolution in all 3 dimensions\cite{bolotnikov2014spectralcorrection}.
Compared to pixelated CZT, these virtual Frisch-grid detectors can be made thicker, require fewer electronic channels, and can utilize lower quality detector material.
Since these virtual Frish-grid detectors have been developed at BNL for ground-based research, there are a number of design considerations necessary for a space-based Compton-telescope application, such as detector packaging and high-voltage risks.  
The AMEGO team is currently developing and optimizing these detectors for space applications and this work is discussed further in Sec.~\ref{sec:ComPairCZT}.
 


The AMEGO Low Energy Calorimeter design uses $8 \times 8 \times 40$~mm virtual Frisch-grid CZT bars. 
The bars are packed in a $4\times4$ array, which share high-voltage and a readout ASIC, as shown in Fig.~\ref{fig:AMEGOCZT}.
This base array, which is $3.7\times3.7\times6$~cm, allows for a modular design that is  readily integrated into large-area arrays.
The Low Energy Calorimeter is segmented into four modules, as can be seen in Fig.~\ref{fig:amego}, each consisting of $5\times10$ CZT base arrays plugged into a motherboard. These modules surround approximately the bottom one-third of the Tracker; two modules are placed below the active area of the Tracker tower and two are rotated to cover each corner of the instrument quadrants, as is seen in Fig.~\ref{fig:amego}. 

The mechanical design for the Low Energy Calorimeter must minimize passive material, much like the Tracker. 
The CZT base array is supported by an interlocking structure fabricated of thin printed circuit board (PCB) material, which, in addition to mechanical support, provides electrical connections for the individual CZT bars. 
Small springs embedded in the walls of this structure guarantee a connection with the side sensing electrodes, which can be seen on the side walls in Fig.~\ref{fig:AMEGOCZT}.
The support structure and CZT module design is based on significant developments made with a prototype detector, which is described further in Sec.~\ref{sec:ComPairCZT}. 

The anode and cathode signals are unipolar; however, the induced signal on the side sensing electrodes is bipolar and wave-front sampling is required to get an accurate measure of the pad signals to determine the 3D interaction position within the detector. 
The 16 bars in the CZT base array, with 6 channels per bar, are read out with a single IDEAS IDE3421 digital ASIC located on the anode side of the array. 
The cathode and pad signals are routed through the side walls of the PCB array structure. 
The 4~kV high voltage required to bias these 4~cm thick detectors is distributed on the top detector board, as seen in Fig.~\ref{fig:AMEGOCZT}.
Each module plugs into the motherboard, which has a control FPGA for each CZT module (5$\times$10 CZT base arrays).

The CZT Low Energy Calorimeter is a novel subsystem which enhances AMEGO's low-energy response; however, many of the components have flight heritage.
CZT pixel detectors have flown on \textit{Swift}-BAT\cite{bat}, AstroSat~\cite{astrosat}, NuSTAR~\cite{nustar}; however, the virtual Frisch-grid detectors have not yet been space qualified. 
The commercially available IDE3421 ASIC is from a family of IDEAS ASICs with flight heritage. 
We asses the TRL of the Low Energy Calorimeter to be 4, which is the lowest TRL system in AMEGO.

\subsection{High Energy Calorimeter}
\label{sec:AMEGOCsI}

To extend the sensitivity up to 1~GeV, the High Energy Calorimeter must provide approximately five radiation lengths to fully contain the  highest energy pair-conversion events.
Many of the subsystem requirements can be extracted from the \fermi-LAT calorimeter, since it has the same function; however, the AMEGO calorimeter must be optimized for lower energy pair events that extend beyond the Low Energy Calorimeter.
This low-energy optimization can be achieved by lowering the energy threshold and improving the energy resolution of the calorimeter.
A modest position resolution is needed ($\sim$cm) to image the large electromagnetic showers that are created as the pair products go through the calorimeter.
The AMEGO High Energy Calorimeter has been optimized to extend the energy response of AMEGO beyond $\sim$100~MeV.

\begin{figure}[b]
\centering
\begin{minipage}{0.48\textwidth}
    \centering
    \includegraphics[height=2.3in]{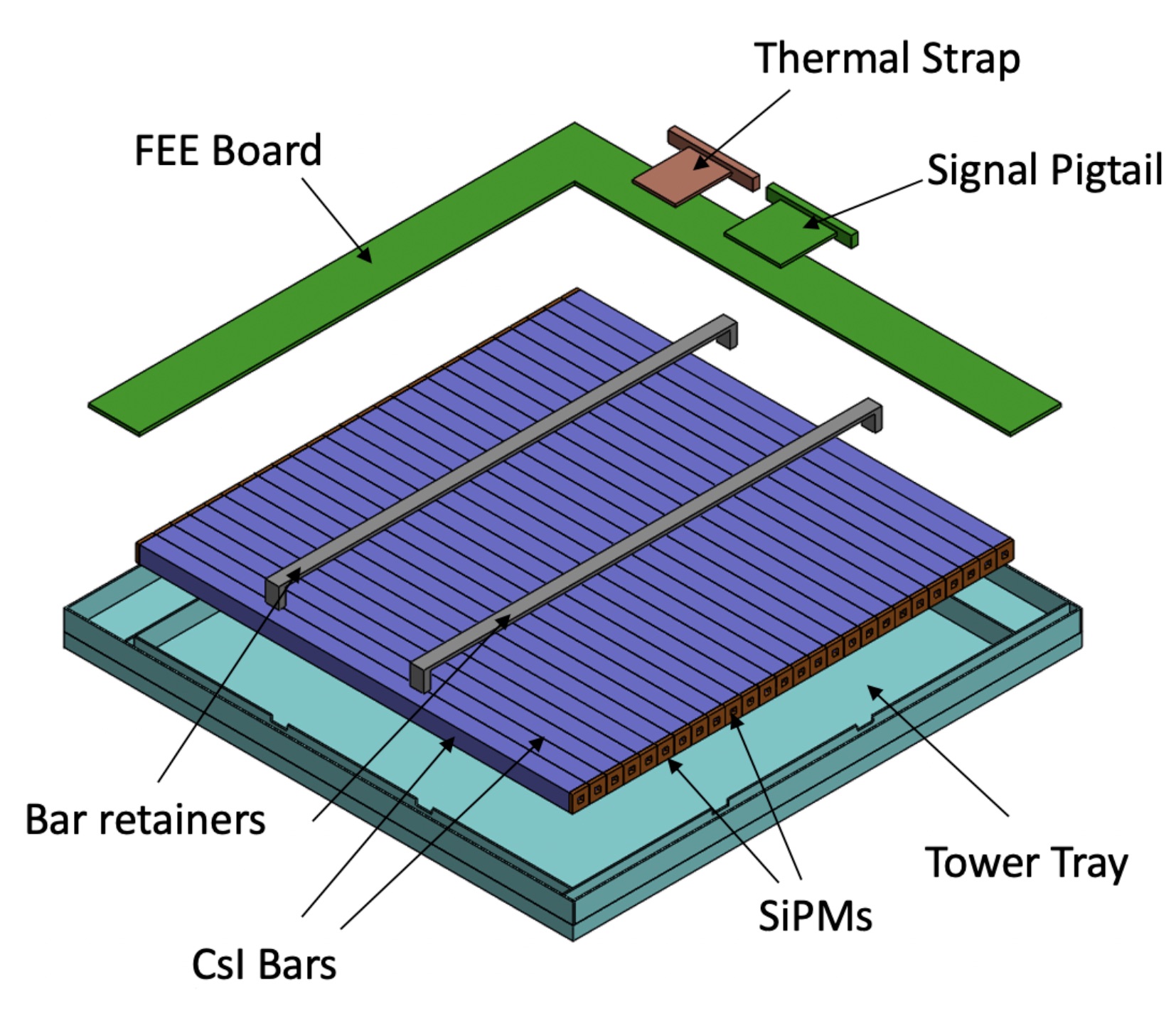}
    \caption{A single layer of the AMEGO High Energy Calorimeter has 26 $1.5\times1.5\times38$~cm CsI bars. The scintillation light is collected by SiPMs on either end, and the signals are fed to the FEE boards on the outer edges of the layer. The bars are supported by a composite tower tray and bar retainers.}
    \label{fig:AMEGOCsI}
\end{minipage}%
\hfill
\begin{minipage}{0.48\textwidth}
    \centering
    \includegraphics[height=2.3in]{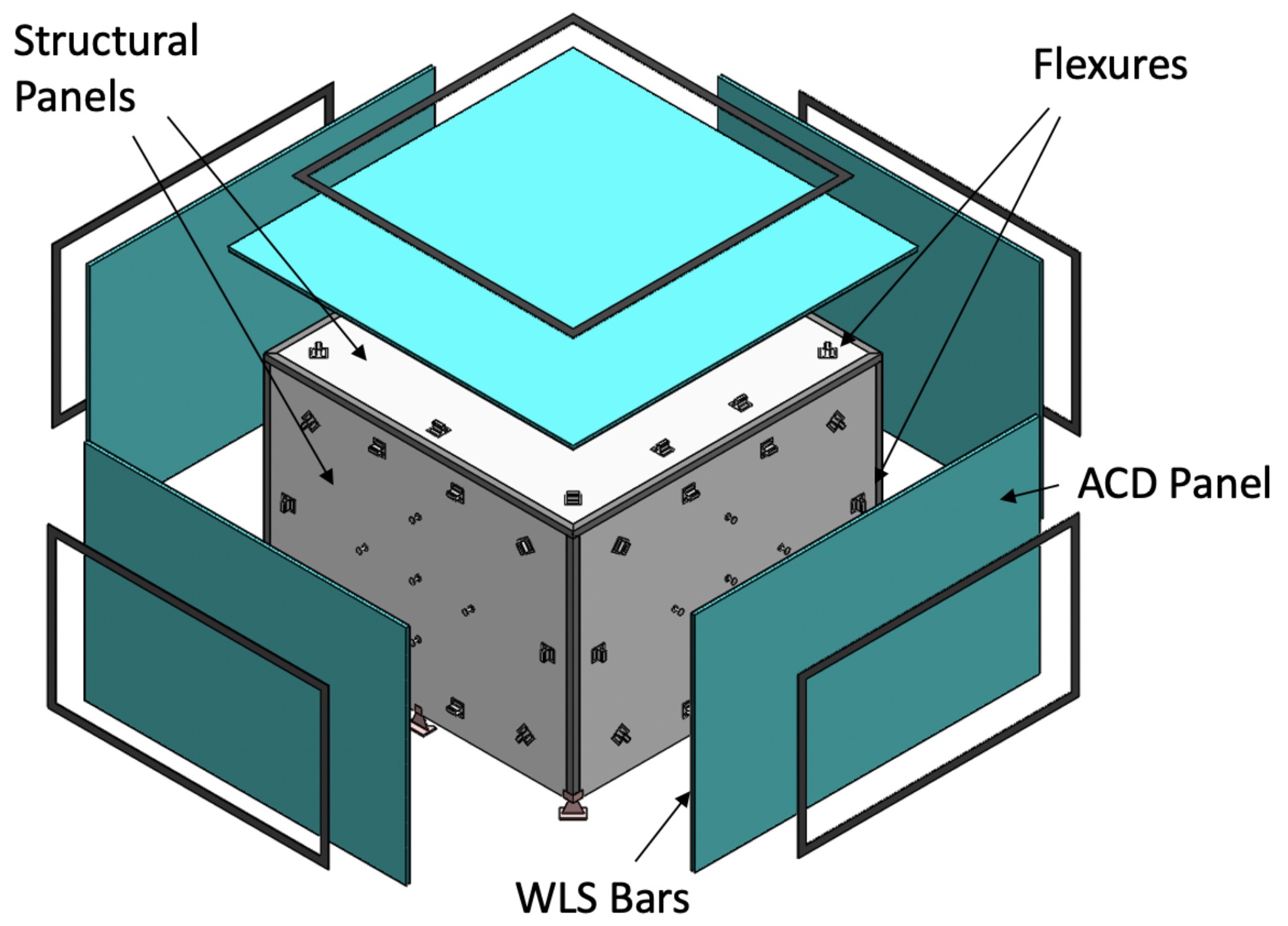}
    \caption{The AMEGO ACD consists of five plastic scintillator panels with wavelength shifting (WLS) bars on each edge. The WLS bars are read out by SiPMs. Each panel is supported by the composite sturctural panels and frame, as shown in the exploded view of the ACD subsystem here.}
    \label{fig:AMEGOACD}
\end{minipage}
\end{figure}

The AMEGO High Energy Calorimeter design is modeled after the \fermi-LAT calorimeter and uses CsI scintillators with SiPM readout to enhance the low energy response of the calorimeter.
Thallium doped CsI bars that are each $1.5\times1.5\times38$~cm are arranged in 6 layers in a hodoscopic pattern, with each alternating layer oriented orthogonal to the one above. 
A single layer of the High Energy Calorimeter contains 26 bars to cover the total area beneath the Tracker, as shown in Fig.~\ref{fig:AMEGOCsI}.
The bars are wrapped in reflective material to maximize light collection efficiency. 

Each High Energy Calorimeter layer of 26 bars is supported in a tray made of composite materials. 
While passive material can still have adverse affects, it is not as critical in the High Energy Calorimeter since the energy resolution requirements for high energy pair events is not strict.
However, low-Z material is still important to reduce the activation of the instrument.

The CsI scintillation light is collected on both ends of the CsI bars by an array of SensL-J series SiPMs. The energy of the interaction is given by the geometric mean of the pedestal-subtracted pulse heights of the two signals. The relative amplitude of signals on each end determines the depth of interaction along the bar to $\sigma = 1$~cm resolution at 1~MeV.
The SiPM signals are processed with IDEAS VA32TA6 ASICs located on the FEE boards at the perimeter of the layer, and 
the corner of each layer contains a digital backend board with an FPGA that controls the ASICs for a single layer and communication with the rest of the instrument
With the use of the SiPM readout, the AMEGO calorimeter has significant improvement over \fermi-LAT at low energies 
with a 
threshold of 60 keV, compared to 600~keV for the \fermi-LAT PIN diode readout~\cite{Woolf2018}.

The CsI High Energy Calorimeter has been modeled off of the \fermi-LAT calorimeter and is being designed and built by the same team at the Naval Research Laboratory (NRL). 
The \fermi-LAT has given hodoscopic arrays of CsI scintillating bars flight heritage. 
The SensL-J SiPMs used in the AMEGO readout have been recently flight qualified on SIRI\cite{siri}, and will soon be flown on BurstCube \cite{burstcube}.
The IDEAS VA32TA6 ASIC have flight heritage on Astro-H\cite{astrohHXI}, eXTP\cite{extp}, and CALET\cite{calet}.
We asses the TRL of the High Energy Calorimeter subsystem to be 6.

\subsection{Anti-Coincidence Detector}
\label{sec:AMEGOACD}

The dominant source of background for gamma-ray telescopes in orbit comes from the abundance of cosmic rays. The shear number of cosmic rays is orders of magnitude more than gamma rays from astrophysical sources, and thus an effective rejection of these background events is required to keep event rates down and not overwhelm the system with unwanted data.
All modern high-energy telescopes use an ACD to reject the cosmic-ray background, and the subsystem design is well understood. 
For example, the \fermi-LAT plastic scintillator ACD surrounds the LAT Tracker and is segmented to avoid self-vetoing from the 
backsplash in high-energy pair events. As AMEGO is optimized for lower energies, this level of segmentation is not necessary.


The AMEGO ACD consists of five plastic scintillator panels that surround the Tracker and Low Energy Calorimeter.
The ACD does not extend to the sides of the High Energy Calorimeter to avoid self-vetoes from the electromagnetic shower in pair events.
The AMEGO ACD side panels are $134\times87\times1.5$~cm, while the top panel is 134 square cm. 
Wavelength shifting bars are inserted in each edge of each panel to increase the light collection uniformity.
The ACD plastic scintillator panels are supported by a low-Z composite structure panels with flexures, as shown in Fig.~\ref{fig:AMEGOACD}.


The ACD wavelength shifting bars are read out on each end by SensL-J series SiPMs, the same SiPMs that are being used for the High Energy Calorimeter. 
No internal interaction position information is attained as the only requirement for the ACD is to effectively veto charged-particles.
A VATA64HDR16 ASIC is used for the readout of each panel, and a single FPGA controls the full ACD.

The AMEGO ACD is closely modeled after the \fermi-LAT ACD; however, single panels of scintillating plastic are used instead of the segmented ACD of the LAT. 
The SensL-J series SiPMs, which are also being used for the High Energy Calorimeter readout, have been recently flight qualified on SIRI\cite{siri}.
The VATA64HDR16 ASIC is part of a family of ASICs that has flight heritage on Astro-H\cite{astrohHXI}, eXTP\cite{extp} and CALET\cite{calet}.
We have assessed the AMEGO ACD to be TRL 6.

\subsection{Observatory Operations}

The AMEGO mission will fly in a low-inclination (6$^{\circ}$), low-earth orbit (600~km) and the prime mission will be 5 years. 
The low-inclination angle is important to minimize the transit through the South Atlantic Anomaly (SAA) which increases the background due to activation of the instrument. 
With its wide FOV, AMEGO operates predominantly in a survey mode, scanning the full sky every 3 hours (2 orbits). 
AMEGO will also have the capabilities to perform inertial target pointing for targets of particular science interest.

With an estimated data volume of 45~GB per day, AMEGO will use a high bandwidth Ka-band communications subsystem for data downlink. 
The NASA Space Network will provide the primary space-to-ground link, and with a 5~min TDRSS contact every orbit, AMEGO has significant margin on the downlink capability.

\section{EXPECTED PERFORMANCE}
\label{sec:performance}

We have performed detailed simulations to study the expected performance of AMEGO using the Medium Energy Gamma-ray Astronomy Library (MEGAlib)~\cite{2006NewAR..50..629Z} software package. 
MEGAlib can perform Monte-Carlo simulations of an instrument's response to point sources and the background radiation environment  in orbit.
An accurate mass model of AMEGO, which includes all detector material and properties, as well as an approximation for the passive material, has been built to determine a realistic performance.
Simulations were done to determine the energy and angular resolution, the effective area, and ultimately the sensitivity to sources of continuum and narrow-line emission.

\begin{figure}[tb]
\begin{minipage}{0.5\textwidth}
    \centering
    \subcaptionbox{Effective Area}{\includegraphics[width=0.9\textwidth]{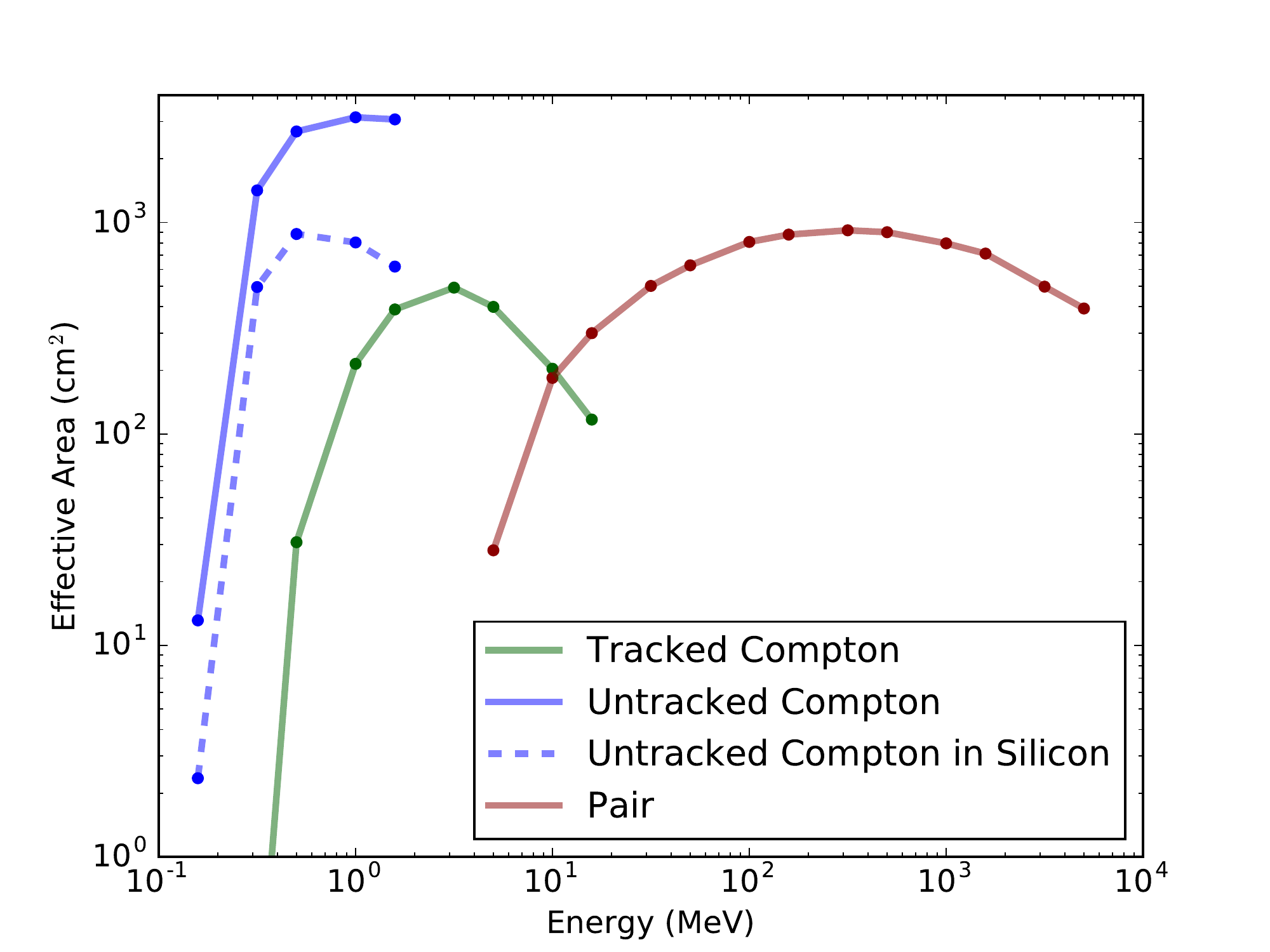}}
\end{minipage}%
\hfill
\begin{minipage}{0.5\textwidth}
    \centering
    \subcaptionbox{Angular Resolution}{\includegraphics[width=0.9\textwidth]{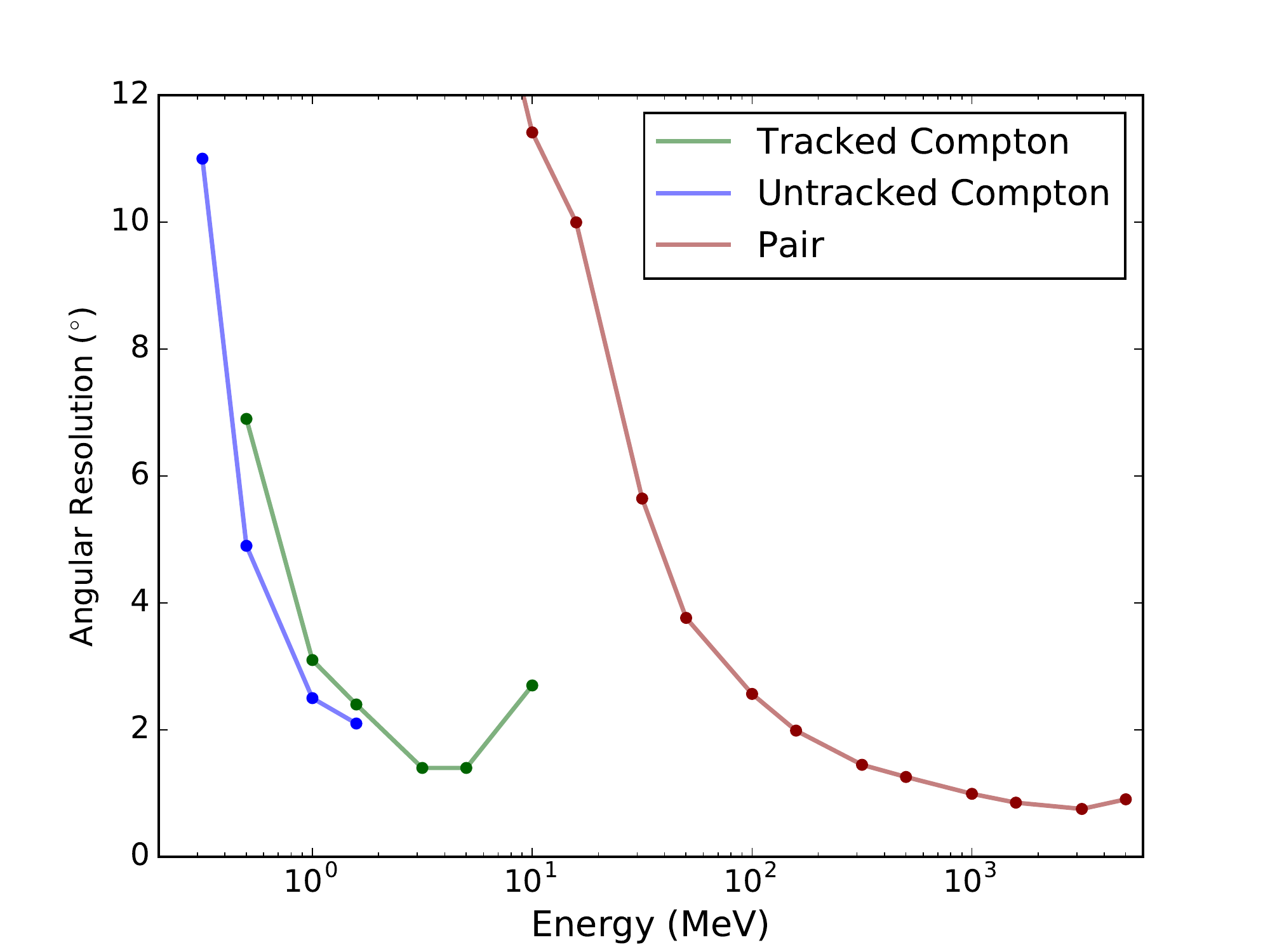}}
\end{minipage}%
\hfill
\begin{minipage}{0.5\textwidth}
    \centering
    \subcaptionbox{Energy Resolution}{\includegraphics[width=0.9\textwidth]{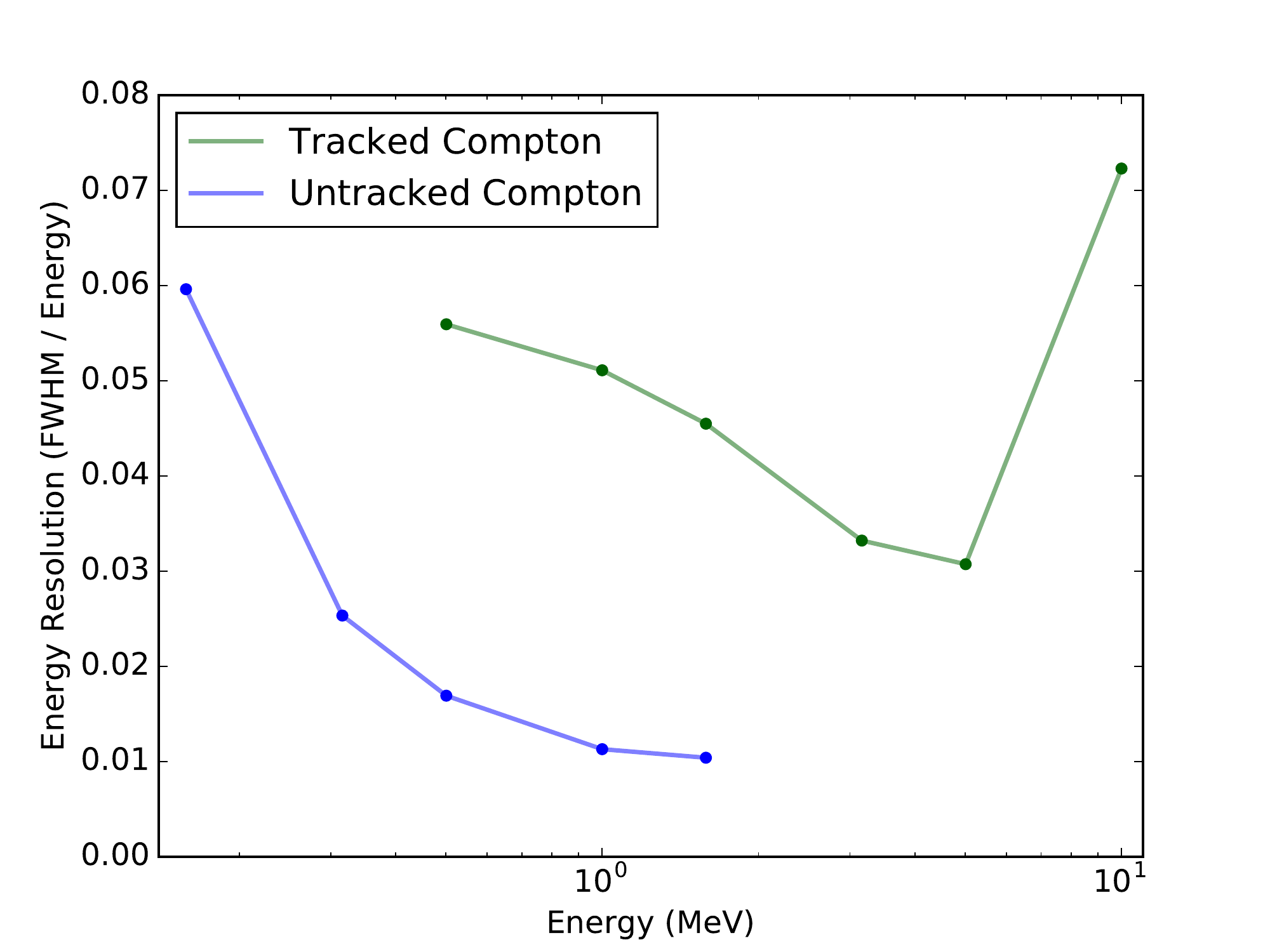}}
\end{minipage}%
\hfill
\begin{minipage}{0.5\textwidth}
    \caption{(a) An effective area of 500--1000~cm$^{2}$ is achieved by combining three different event classifications across four orders of magnitude. 
    (b) For  Compton and pair events, the angular resolution is 2.5$^{\circ}$ at 1~MeV and 1$^{\circ}$ at 1~GeV, respectively.
    (c) An energy resolution of 1\% FWHM/E is achieved at 1~MeV. See the text for further discussion of each instrument parameter.}
    \label{fig:effectivearea}
\end{minipage}
\end{figure}

To better understand the telescope response, we define three event classifications for AMEGO:
\begin{itemize}
    \item \textit{Untracked Compton events}, generally with the lowest incident energy ($\lesssim 1$ MeV), are events where the first interaction is a Compton scatter and the Compton-scattered, or recoil, electron is absorbed in a single Tracker layer. The scattered electron's direction can not be measured, and thus is referred to as ``untracked.'' Alternatively, Compton events that interact solely in the Low Energy Calorimeter are also untracked Compton events since the calorimeter has no ability to track the direction of the scattered electron.
    \item \textit{Tracked Compton events}, generally with an intermediate incident energy  ($\sim$~1--10~MeV), are events where the Compton-scattered electron passes through more than one layer of the Tracker and its direction can be measured. This additional kinematic information reduces the known origin of the photon from a circle on the sky to an arc, thereby reducing the background in an observation.
    \item \textit{Pair events}, generally with the highest incident energy ($\gtrsim 10$ MeV), are events where the first interaction is the conversion a gamma ray into an electron-positron pair.
\end{itemize}
We look at each of these event classifications separately for the instrument performance parameters, but the sensitivity calculations use a combined response when more than one event classification is relevant.
Figure~\ref{fig:effectivearea} shows the effective area, angular resolution, and energy resolution that were obtained from MEGAlib simulations, where we have simulated mono-energetic point sources at the instrument's zenith for a sampling of energies across the AMEGO band.
These results will be discussed in detail in the following section prior to the presentation and discussion of the sensitivity plots.

It is worth noting that MEGAlib was first developed for the Medium Energy Gamma-Ray Astronomy (MEGA) instrument built by the Max Plank Institute in the early 2000's~\cite{mega}, which had a very similar design to AMEGO.
Over the past twenty years, MEGAlib has been further developed through work with the COSI collaboration~\cite{cosi} with a focus on the Compton regime.
The MEGAlib pair event identification and reconstruction tools do not yet take into account the state-of-the-art algorithms developed for the \fermi-LAT, and thus we can expect an improvement in the pair regime relative to what is shown here.

\subsection{Effective Area}
The effective area is a measure of the instrument's detection efficiency. 
It is given in units of cm$^{2}$ since it is defined as the area that an \textit{ideal} absorber needs to detect an equivalent number of incident photons.
Here, we defined the effective area as the number of valid events qualified by MEGAlib's reconstruction tool, \textit{revan}, divided by the number of simulated incident events entering the telescope, and scaled by the simulated area surrounding the mass model.
The events are separated based on the three classification defined above and there is no further selection on the reconstructed energy or photon origin.

Figure~\ref{fig:effectivearea}~(a) shows the simulated effective area for the three event classifications in AMEGO.
By combining the detection of Compton and pair events, AMEGO has an effective area that ranges from 500 to 1000~cm$^{2}$ across four decades of energy.
The untracked Compton events are further separated into two categories. 
``Untracked Compton in Silicon'' events require their first interaction to be in the silicon Tracker. 
If this restriction is not made and we also allow events which interact first in the CZT Low Energy Calorimeter, the effective area increases dramatically. 
Although the Low Energy Calorimeter cannot track the direction of the scattered electron, with the excellent energy and position resolution of the CZT, these events can still be reconstructed and used for imaging, particularly for sources of narrow-line emission. 
The effective area is not directly an instrument requirement, but it feeds into the sensitivity, as discussion in Sec.~\ref{sec:AMEGOsensitivity}.

\subsection{Angular Resolution}
The angular resolution is defined in different ways for Compton and pair events.
The angular resolution for a Compton telescope is described by the FWHM of the 1D point spread function (PSF), also known as the ARM distribution. The ARM, or angular resolution measure, is the smallest angular distance between the nominal source location and the event circle for each event; the distribution of all ARM values from a sample of Compton events gives effective 1D PSF of a Compton telescope.
In the pair regime, the angular resolution is defined as the 68\% containment radius of the PSF. The pair PSF is given by the angular distance between the nominal and reconstructed photon direction for events for a point source.

Figure~\ref{fig:effectivearea}~(b) shows the simulated angular resolution for the three event classifications in AMEGO.
An angular resolution of 2.5$^{\circ}$ at 1~MeV and 1$^{\circ}$ at 1~GeV, as demonstrated here, satisfy the requirements needed to address AMEGO's science goals.

Measuring the track of a Compton-scattered electron is not expected improve the ARM for an event since the direction of the Compton-scattered electron is so poorly measured. 
In fact, as shown in Fig.~\ref{fig:effectivearea}~(b), the angular resolution for tracked events is slightly worse than untracked events for the same energies. 
This, however, is due to a selection effect: for an event to be tracked at 1~MeV, for example, a significant fraction of energy must be transferred to the Compton-scattered electron, which, by the Compton equation, will result in a large Compton scatter angle.
Events with a larger Compton scatter angle inherently have a worse angular resolution due to the propagation of error in the Compton equation.

\subsection{Energy Resolution}
The energy resolution is given by the FWHM of the reconstructed photopeak reported as a percentage of the incident energy $\Delta E/E$. 
The energy resolution for pair events, which was found to be $\sim$10\% at 1~GeV, is not shown here since it is not an instrument requirement. 
However, as discussed above, we expect the energy resolution in the pair regime to improve once the \fermi-LAT reconstruction tools are implemented.

Figure~\ref{fig:effectivearea}~(c) shows the energy resolution for Compton events.
An energy resolution of 1\% FWHM/E is achieved at 1~MeV. 
The energy resolution for Untracked Compton events is better than that seen for tracked Compton events for two reasons. First, the Low Energy Calorimeter dominates the Untracked Compton event classification and the CZT has better energy resolution than the DSSDs in the Tracker. Second, the energy resolution for tracked events will often be worse since more interactions are recorded (at least two in the tracker, by definition), and the errors add up for each measurement.

\subsection{Continuum and Narrow-Line Sensitivity}
\label{sec:AMEGOsensitivity}

\begin{figure}[tb]
    \centering
    \includegraphics[height=2.2in]{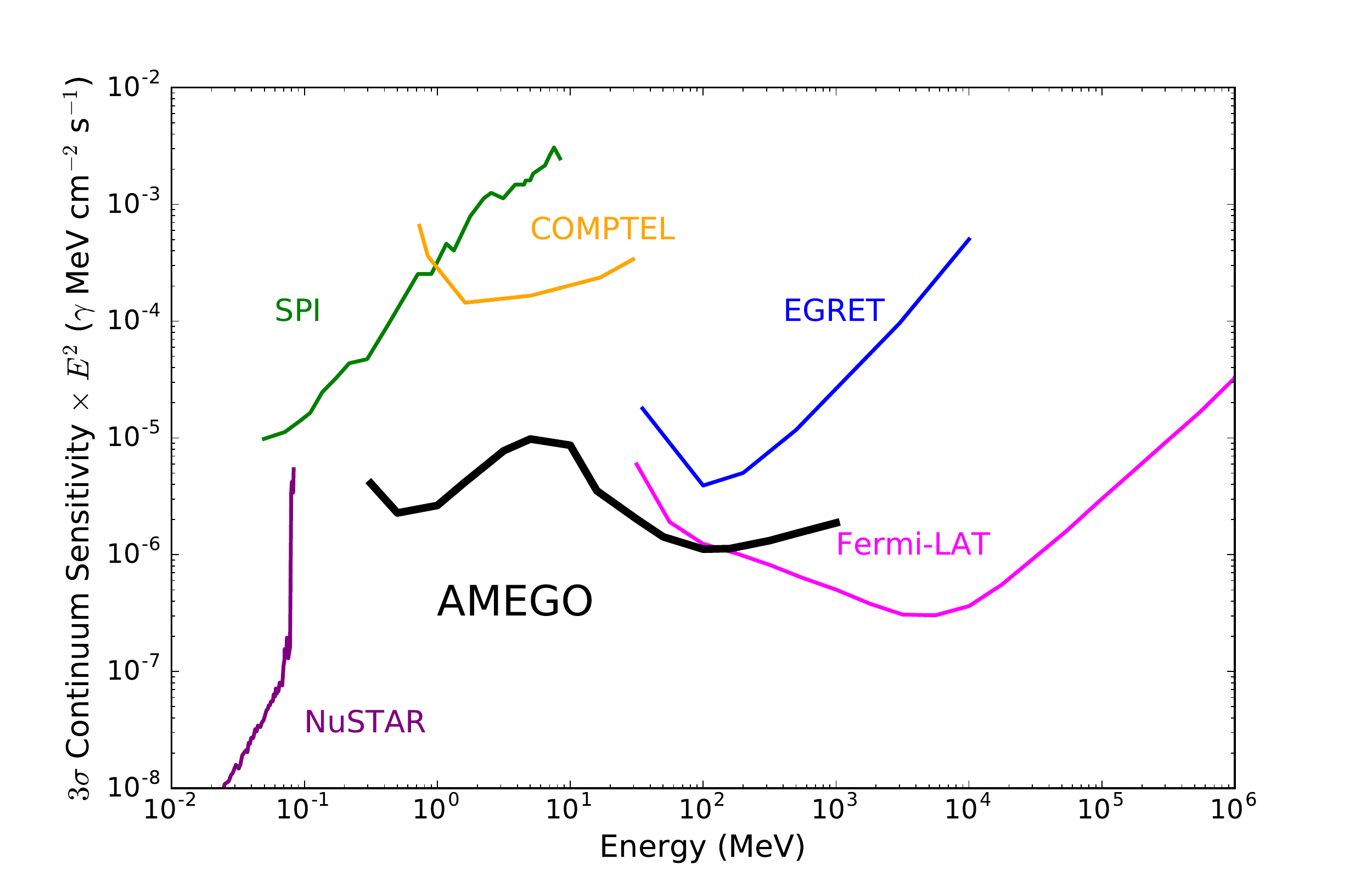}
    \includegraphics[height=2.2in]{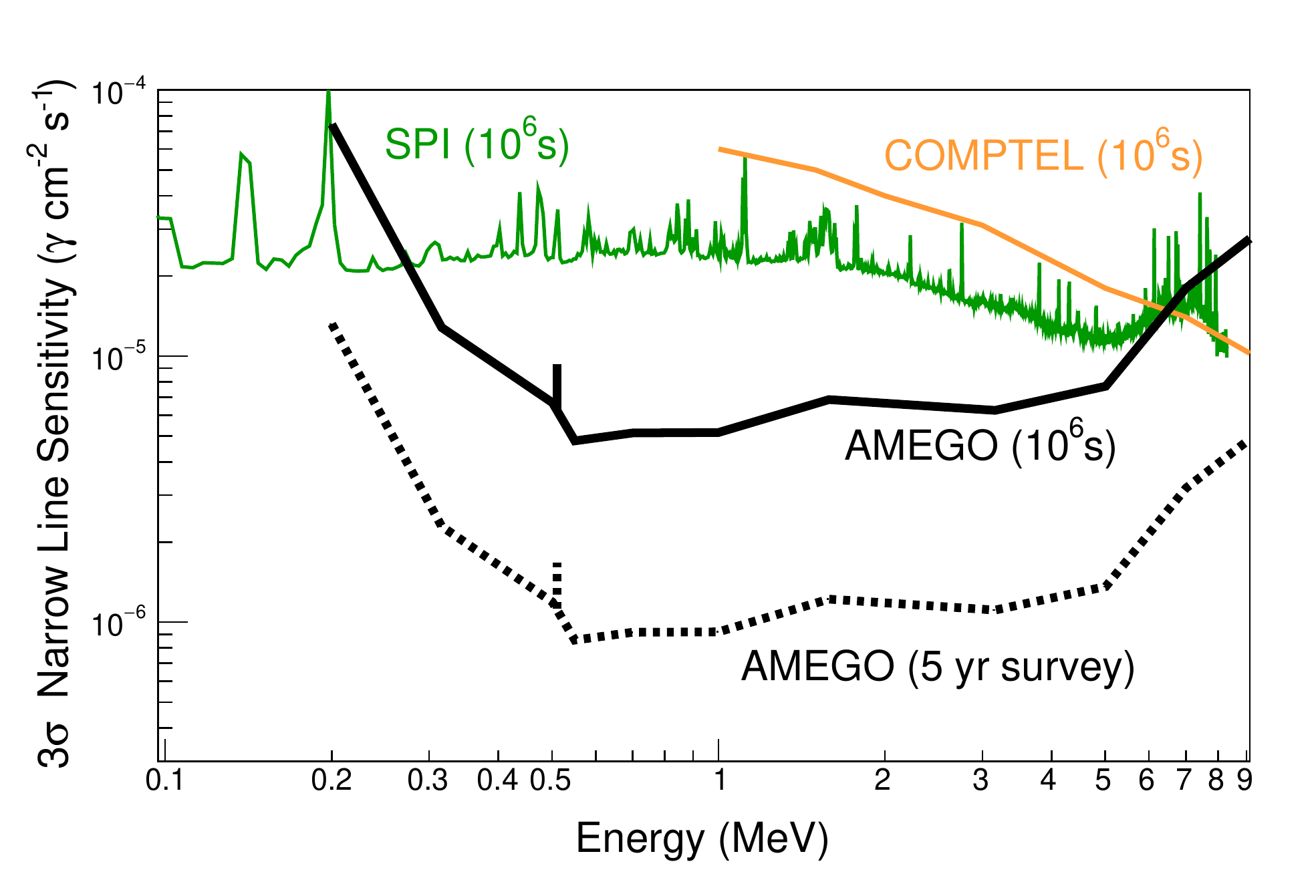}
    \caption{\textit{Left} The $3 \sigma$ on-axis point source continuum sensitivity for a 5 year AMEGO mission compared with the \fermi-LAT (same incident angle and efficiency over 5 years), COMPTEL\cite{Comptel:1993} and EGRET\cite{egret} (40\% efficiency over two weeks), and
    NuSTAR\cite{nustar} and SPI\cite{spi} (exposure of 10$^{6}$~seconds). We assumed a 5-year mission with a 20\% observation efficiency (due to field of view and South Atlantic Anomaly).
    \textit{Right} The 3$\sigma$ narrow-line sensitivity for AMEGO is compared to INTEGRAL/SPI and COMPTEL.
}
    \label{fig:sensitivity}
\end{figure}

The sensitivity of a telescope is a measure of its capability to detect faint a sources; a lower sensitivity is better.
For gamma-ray telescope, the sensitivity can be calculated based on the background rate, the effective area, the angular resolution, and, in the case of the narrow-line sensitivity, the energy resolution. 

The sensitivity has been calculated differently for the two regimes of the AMEGO telescope. In the Compton regime ($\lesssim$10~MeV), where the background is dominated by activation in the instrument and surrounding passive material, we have performed full background simulations in MEGAlib which include activation. 
We have then used MEGAlib's \textit{SensitivityOptimizer} program to determine the continuum sensitivity for this range.
In the pair regime ($\gtrsim$10~MeV), where the backgrounds are well understood and modeled from \fermi-LAT observations, we have calculated the sensitivity analytically by
\begin{equation}
    I_{src} = \frac{E}{A_{eff} T_{obs}} \times \left( \frac{n_{sig}^2}{2} + \sqrt{ \frac{n_{sig}^4}{4} + \frac{n_{sig}^2 A_{Eff} T_{obs} N_B d\Omega}{E}} \right),
\end{equation}
where $E$ is the energy, $A_{eff}$ is the effective area, $T_{obs}$ is the observation time, $n_{sig}$ is the significance (3$\sigma$ is used here), and $N_B$ is the background. 
The parameter $d \Omega$ is defined as $2\pi (1-\cos (2\times PSF))$, with $PSF$ given by the angular resolution.
The background models used for both the input to the low energy MEGAlib simulations and the high energy analytical calculation include Galactic, extra-galactic, and diffuse emission, while the activation simulations also include models of cosmic-ray particles in low-earth orbit. 

Figure~\ref{fig:sensitivity} \textit{left} shows the continuum sensitivity for the AMEGO five year mission compared to other x-ray and gamma-ray telescopes in the neighboring energy bands.
AMEGO will provide over an order of magnitude improvement in continuum sensitivity over previous instruments, which will enable the next generation of multi-messenger astrophysics.
With AMEGO's prime operation in survey mode, an observation efficiency of 20\% is assumed for the sensitivity.

The ``W'' shape of the sensitivity curve is a result of the competing interaction processes in AMEGO. 
Where these competing processes overlap $\sim$~10~MeV, the current reconstruction algorithms in MEGAlib struggle with event classification; however, there are current efforts to use machine learning approaches to improve this~\cite{Zoglauer2020}.
We expect this bump to be reduced and a smooth line between the response at 1~MeV and at 100~MeV should be attained. 

The narrow-line sensitivity demonstrates the sensitivity of a telescope to sources of line emission. 
This is done by taking the energy resolution into account in the sensitivity calculation.
Since the requirements for narrow-line emission are limited to the Compton regime, we only calculate the narrow-line sensitivity between 200~keV and 10~MeV. 
For these calculations, we have used the full background simulation which takes into account activation of the instrument and the mono-energetic simulations discussed in the previous section.

Figure~\ref{fig:sensitivity} \textit{right} shows the narrow-line sensitivity for AMEGO. 
A sensitivity of close to an order of magnitude is achieved relative to INTEGRAL/SPI\cite{spi}, which will allow for progress in resolving the processes of element formation in extreme environments, one of AMEGO's main science goals. It is important to note that SPI has a 16$^{\circ}$ FOV and performs pointed observations, while the AMEGO sensitivity is for all-sky exposures.


\section{HARDWARE PROTOTYPE}
\label{sec:prototype}

\begin{figure}[t]
    \centering
    \includegraphics[height=2.4in]{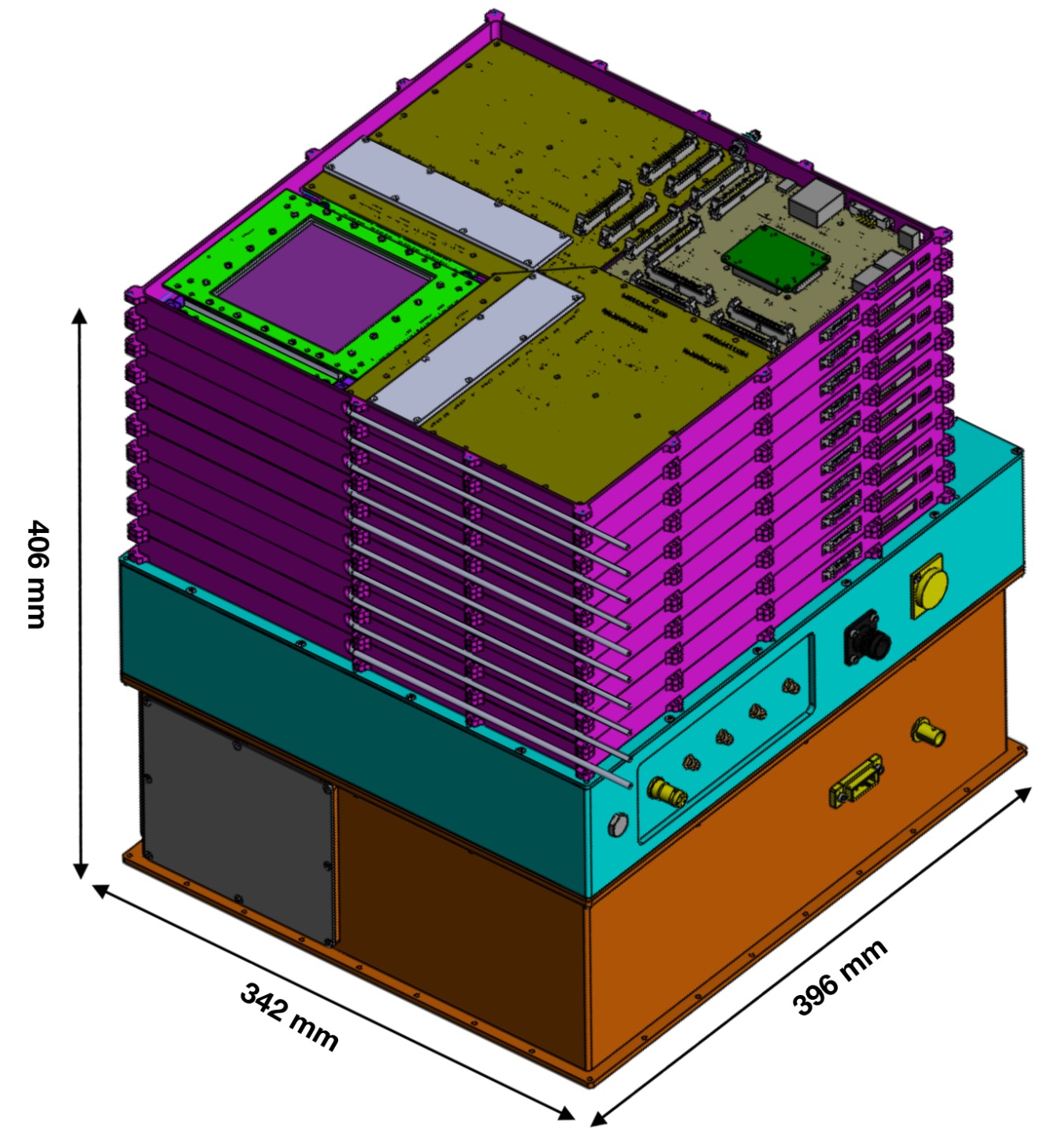}
    \includegraphics[height=2.4in]{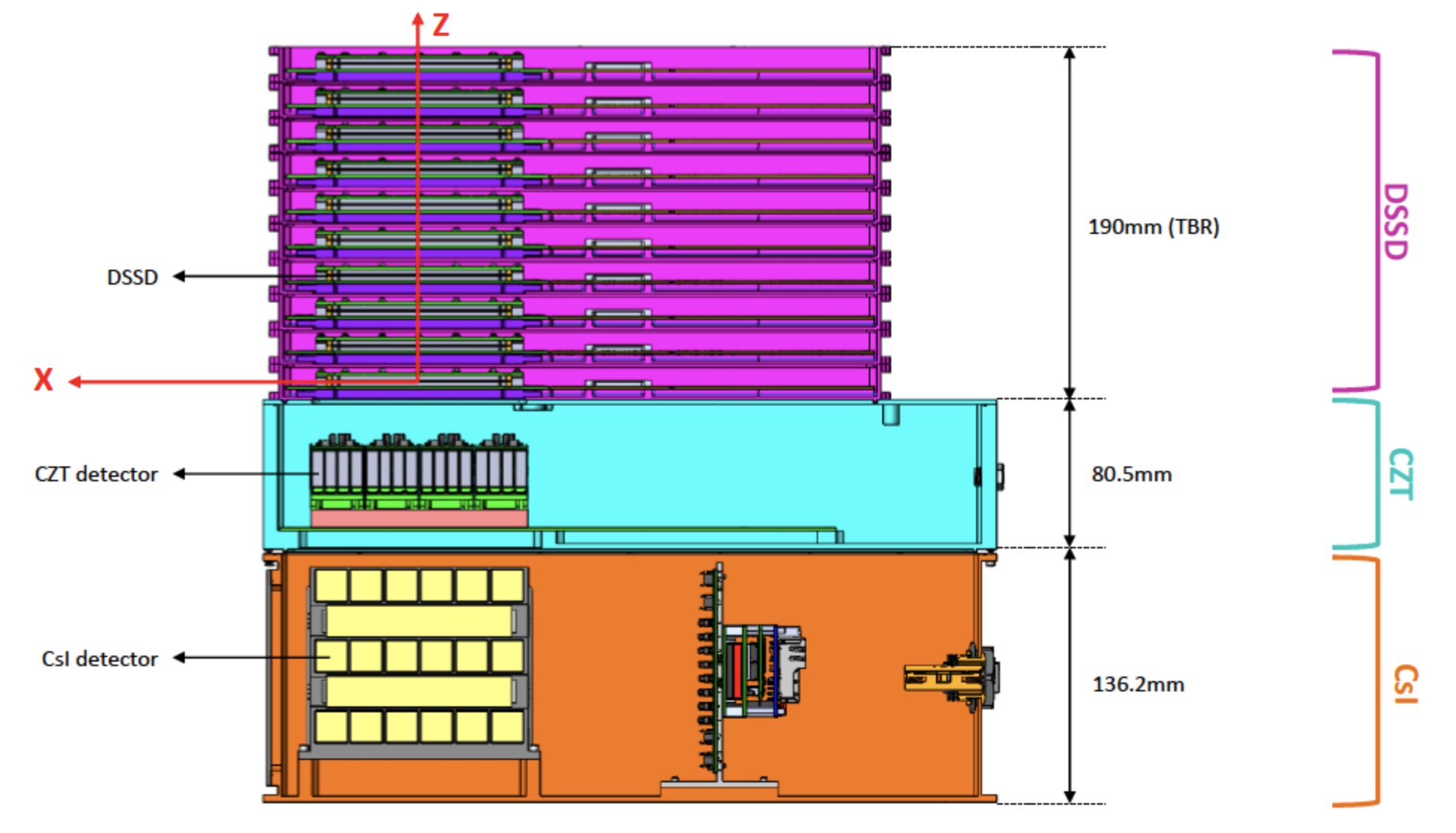}
    \caption{The AMEGO prototype instrument is being built to advance and validate the hardware and software tools to be used in the AMEGO mission. Each prototype detector, housed in its own instrument box, fills a $10 \times 10$~cm$^2$ area and is stacked to minimize the distance between the separate subsystems. The Tracker consists of 10 layers of $10 \times 10$~cm$^2$ DSSDs, each supported and enclosed in an aluminum tray. Directly below the Tracker is a single layer of 2~cm thick virtual Frisch-grid CZT detectors, and then finally, 5 layers of CsI  bars hodoscopically arranged. Not pictured here is the ACD detector which surrounds the full prototype instrument for the technology demonstration balloon flight in 2022.}
    \label{fig:ComPair}
\end{figure}

The AMEGO team has been developing a prototype instrument over the past few years to advance and validate the hardware and software tools used in AMEGO.
The prototype, often referred to as \textit{ComPair} in the literature, is funded through several NASA Astrophysics Research and Analysis (APRA) grants.
The four subsystems of AMEGO are being developed for the prototype instrument, but on a much smaller scale; see Fig.~\ref{fig:ComPair}.
The goal of the AMEGO prototype is to verify the AMEGO subsystems work together to detect and reconstruct both Compton and pair conversion events in a relevant environment. 
The team is working towards a gamma-ray beam test of the prototype instrument at the High Intensity Gamma-ray Source (HIGS) at Duke University, and a technology demonstration balloon flight in Fall 2022 (delayed due to COVID-19) from Fort Sumner, NM. 
An overview of the subsystems for the AMEGO prototype, description and current status, as well as the integration and testing plan is given in the following sections.

\subsection{Si Tracker Prototype}
\label{sec:ComPairTracker}

\begin{figure}[tb]
    \centering
    \includegraphics[height=2.1in]{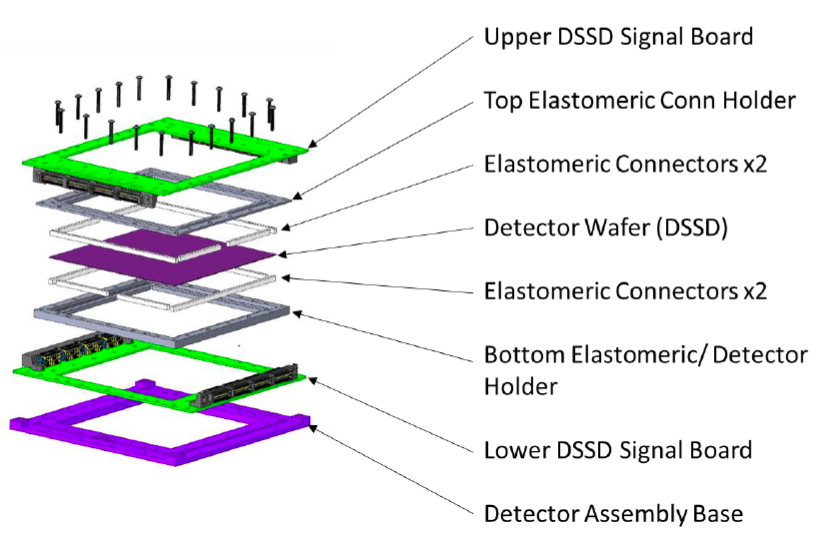}
    \hfill
    \includegraphics[height=2.1in]{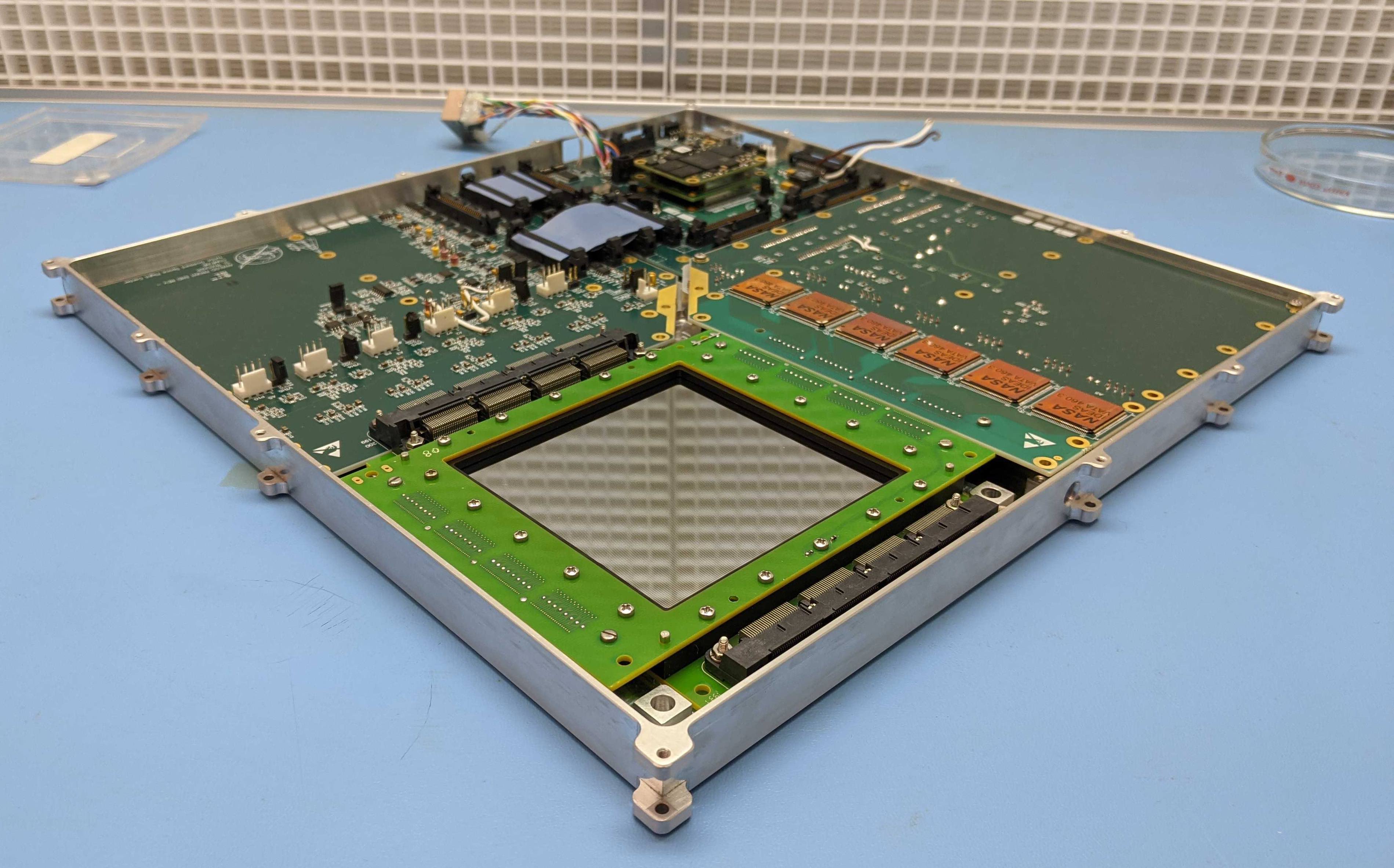}
    \caption{The AMEGO prototype Tracker consists of 10 layers of DSSDs. An exploded view of the custom carrier that uses elastomeric connectors, as opposed to wirebonds, is shown on the left. The photo on the right shows a single fully integrated layer of the Tracker. The DSSD mounted in the custom carrier assembly is seen in the front quadrant, two AFEs which each contain 6 VATA460.3 ASICs are connected to the carrier signal boards (one is flipped upside down so the boards could be identical in design), and the back quadrant houses the DBE which handles the data and communication for the layer.}
    \label{fig:compairtracker}
\end{figure}

The AMEGO prototype Tracker has been designed, built and tested by GSFC, with firmware support from Los Alamos National Laboratory (LANL). 
The subsystem consists of 10 layers of $10\times10$~cm$^2$ DSSDs fabricated by Micron Semiconductor stacked with a 1.9~cm separation. 
The silicon wafers are 500~$\mu$m thick with 510~$\mu$m strip pitch, giving 192 AC-coupled strips per side. 
There have been three revisions of the AMEGO prototype silicon wafers with slight modifications of the strip width and resistor placement.
The GSFC team is currently working with Micron on a fourth revision that aims to optimize the detectors to have low leakage current while minimizing interstrip capacitance so as to achieve the energy resolution goal of the subsystem.  
The wafers have been characterized with bulk leakage current and capacitance measurements, in addition to single strip measurements of the leakage current, interstrip capacitance, resistance, and coupling capacitance.
See Sec.~\ref{sec:AMEGOTracker} for an overview of the Tracker requirements and general operating principle.

The wafers are mounted in a custom carrier, as shown in Fig.~\ref{fig:compairtracker}, which uses elastomeric connections instead of wirebonds to route the signals from each strip to the front-end electronics (FEE). This carrier offers a flexible design, with the option of testing with different readout systems or daisy-chaining detectors to employ the ladder configuration of 4 detectors that the AMEGO design is based on.
Capacitance, and therefore noise, scales as the length of the chained strip, so understanding the performance of the ladder configuration is important for the AMEGO Tracker development. 
The team has also integrated one wire-bonded carrier to understand the possible noise contribution of the elastomeric connector.
The wafers integrated in the custom carriers have been tested with bench-top electronics to confirm functionality.

The custom FEE are based on the IDEAS VATA460.3 ASIC. The ASIC has 32 channels and thus the FEE boards have 6 ASICs per detector side. The analog front end (AFE) boards are designed to handle both positive and negative signals to readout the front (junction side) and back (ohmic side) of the detector. 
A digital back-end board (DBE) with an FPGA handles data from the 12 ASICs, telemetery, and communication with Trigger Module (see Sec.~\ref{sec:ComPairTrigger}) for each layer. Each layer operates entirely independently from the other Tracker layers.
A fully integrated silicon wafer in the custom carrier with the AFE and DBE boards mounted in the aluminum tray is shown in Fig.~\ref{fig:compairtracker}~\textit{right}. 
The AFE and DBE have undergone extensive testing and are ready to mate with the detector. The prototype Tracker, while delayed by COVID-19, is expected to undergo integration and testing in the coming months. See Ref.~\citenum{griffinSPIE} for a more detailed description of the AMEGO prototype Tracker.


\subsection{CZT Low Energy Calorimeter Prototype}
\label{sec:ComPairCZT}

\begin{figure}[tb]
    \centering
    \includegraphics[height=2.0in]{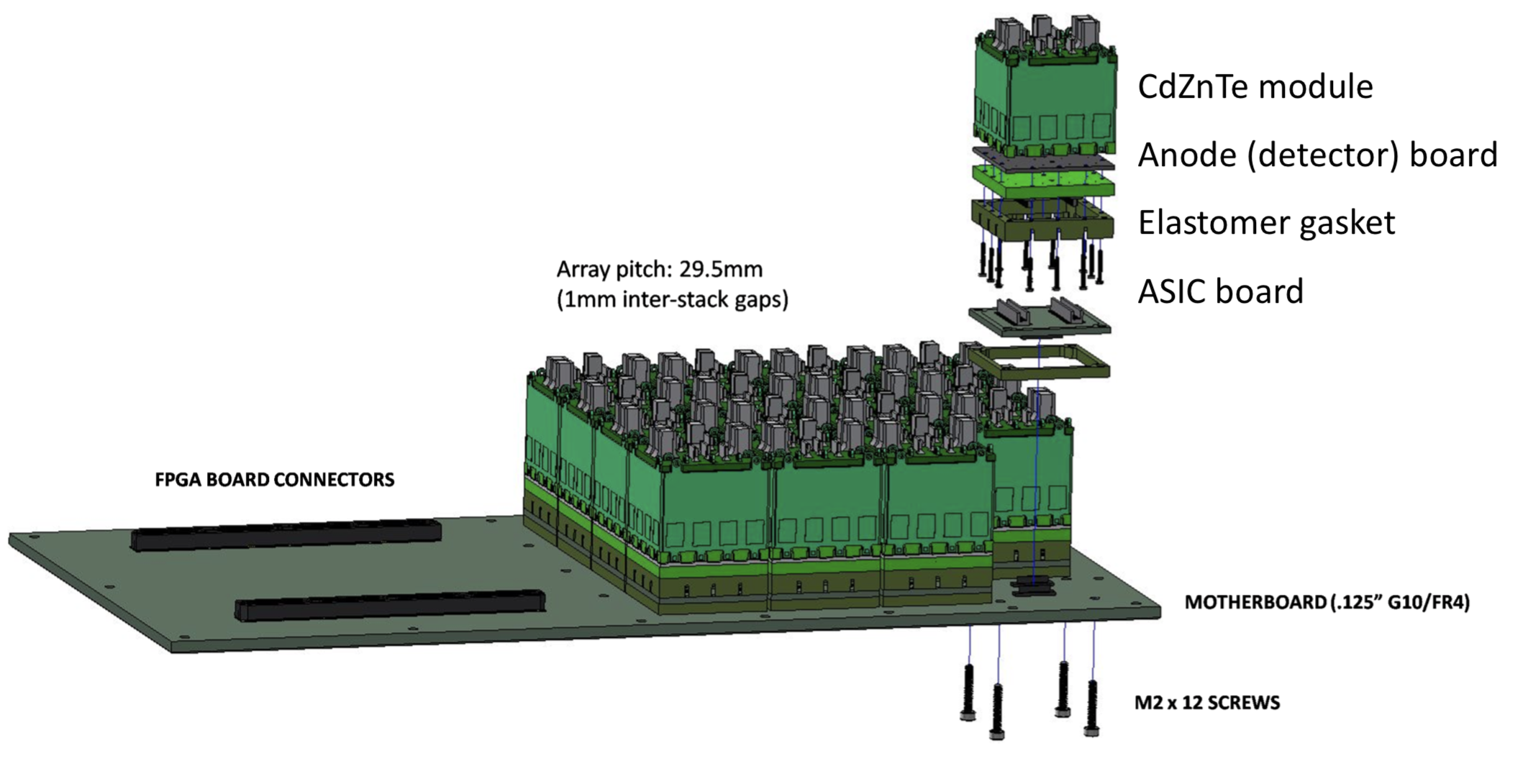}
    \hfill
    \includegraphics[height=2.0in]{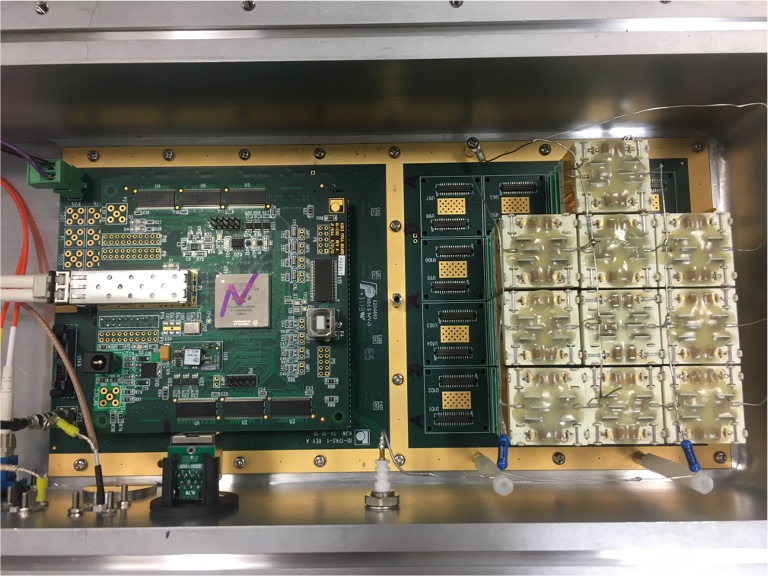}
    \caption{The AMEGO prototype Low Energy Calorimeter consists of 16 modules that are each approximately one cubic inch and contain an array of 4$\times$4 CZT bars. The unique and custom-made module, shown in an exploded view on the left, is made of printed circuit board (PCB) material and connects to a single custom ASIC. The photo on the right shows a recent test of the calorimeter subsystem with 10 integrated CZT detector modules.}
    \label{fig:compairCZT}
\end{figure}

The Low Energy Calorimeter is based on position-sensitive virtual Frisch-grid detectors that have been developed at Brookhaven National Laboratory (BNL)~\cite{Bolotnikov2007}. 
The CZT detectors, which are each 6$\times$6$\times$20~mm$^{3}$, have gold contact on the anode (bottom) and cathode (top), and the relative drift time of the signal gives a measure of the Z coordinate of each interaction.
To make the detector sensitive in three dimensions, four metal pads are placed on the side of the crystal near the anode.
With these virtually grounded pads isolated from the detector, they shield the anode as if a real Frisch-grid were placed inside the detector, and the amplitude of the signals read out from these electrodes are used to evaluate the X and Y coordinates of an interaction. 
Position resolution tests with a pulse laser have found a single crystal position resolution of $\sigma$=0.2-0.3~mm.
Significant groundwork has been completed by the BNL group in developing a 3D voxel correction for the spectral response of these bars, which can achieve excellent energy resolution for single crystals of 0.9\% FWHM at 662 keV~\cite{Bolotnikov2014, bolotnikov2016}. 

The CZT bars are packaged in a module that encloses a 4$\times$4 array that has been designed and fabricated at GSFC using printed circuit board (PCB) material. 
This module, which is 2.5$\times$2.5$\times$2~cm$^{3}$ and is shown in Fig.~\ref{fig:compairCZT}, is the base element of the Low Energy Calorimeter detector plane. 
With each module plugging into a motherboard, one can make an array of any desired size with practically no ``dead space.''
For the AMEGO prototype, the subsystem consists of 16 modules which fill the 10$\times$10~cm$^{2}$ area subtended by the Tracker DSSDs.
The unique custom-made module design has undergone a number of revisions to resolve initial issues with leakage down the edges of the module structure and has since performed well in extensive testing.
In October 2020, a single module was tested with a charged particle beam at the NASA Space Radiation Laboratory at BNL to measure activation and detector polarization. 

The module is read out by a single low-noise custom ASIC designed for the Frisch-grid detector\cite{VERNON20191} that is connected to the anode (bottom) side of the module.
The top board of the module distributes an external 2500~V high-voltage source to bias the detectors.
Figure~\ref{fig:compairCZT} \textit{right} shows 10 modules mounted onto the mother PCB with the FPGA board, shown on the left of the photo, which handles data from all modules, distributes the low-voltage power, and communicates with the Trigger Module.
The CZT Low Energy Calorimeter subsystem with 10 operating modules has undergone extensive testing and calibration.
See Ref.~\citenum{hays2020} for a more detailed description of the AMEGO prototype CZT Low Energy Calorimeter.

\subsection{CsI High Energy Calorimeter Prototype}
\label{sec:ComPairCsI}

The High Energy Calorimeter prototype has been designed, built, and tested at the Naval Research Laboratory (NRL). It consists of 5 layers of six 1.7$\times$1.7$\times$10~cm$^{3}$ CsI bars where each layer is orthogonal to the one above, resulting in a hodoscopic calorimeter. The bars, which are wrapped in Tetratex to increase light collection, are read out on either end by SensL ArrayJ 6$\times$6~mm$^{2}$ quad (2 by 2 array) SiPMs. The relative amplitude of the signals on either end of the CsI bar is used to calculate the depth of interaction along the length of the bar, where the geometric mean of the two signals gives the energy. The prototype bars have a measured position resolution of $\sigma$=1.5~cm at 662 keV and an energy resolution of 3.5\% at 662 keV. 

\begin{figure}[tb]
    \centering
    \includegraphics[height=2in]{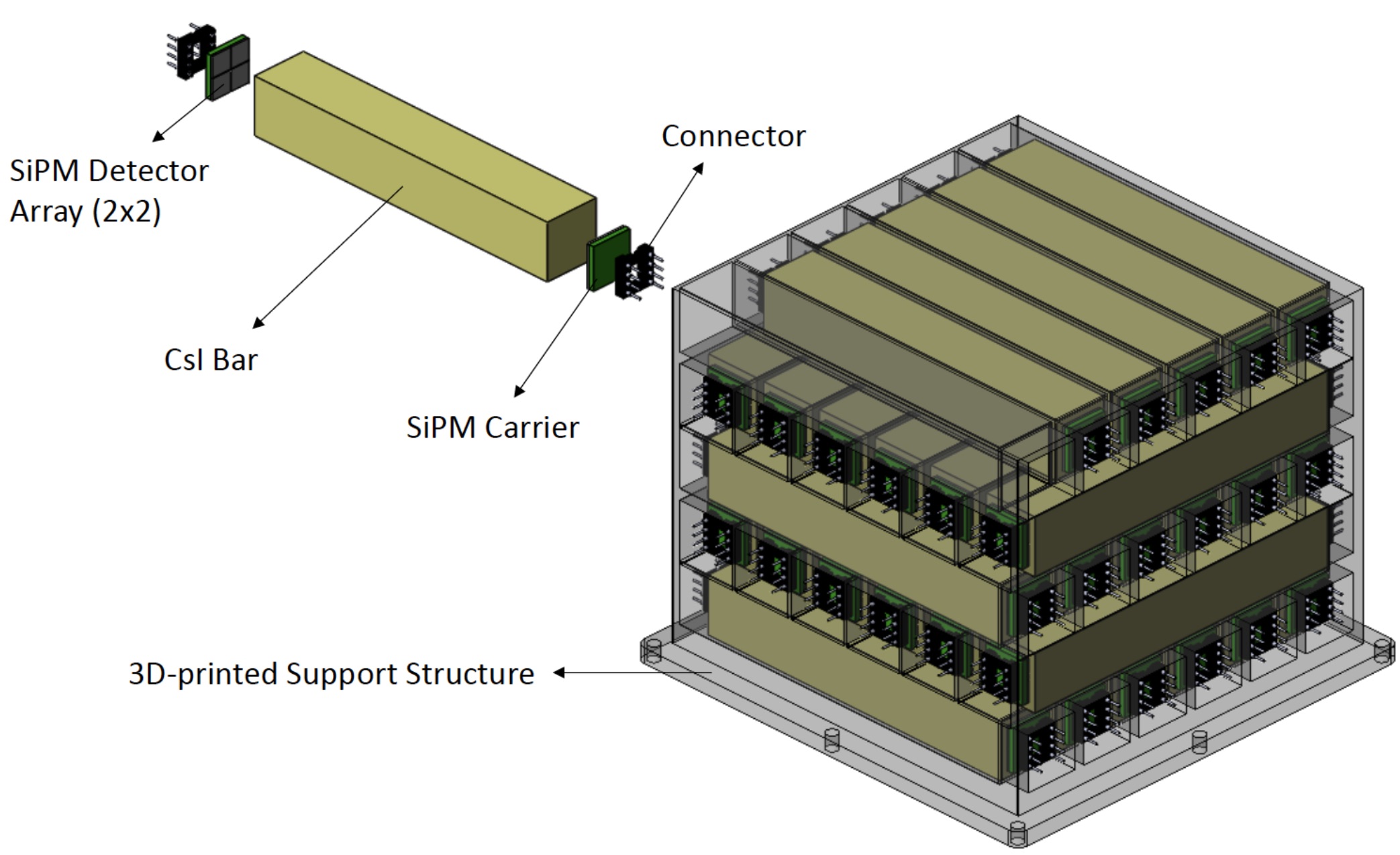}
    \hfill
    \includegraphics[height=2.2in]{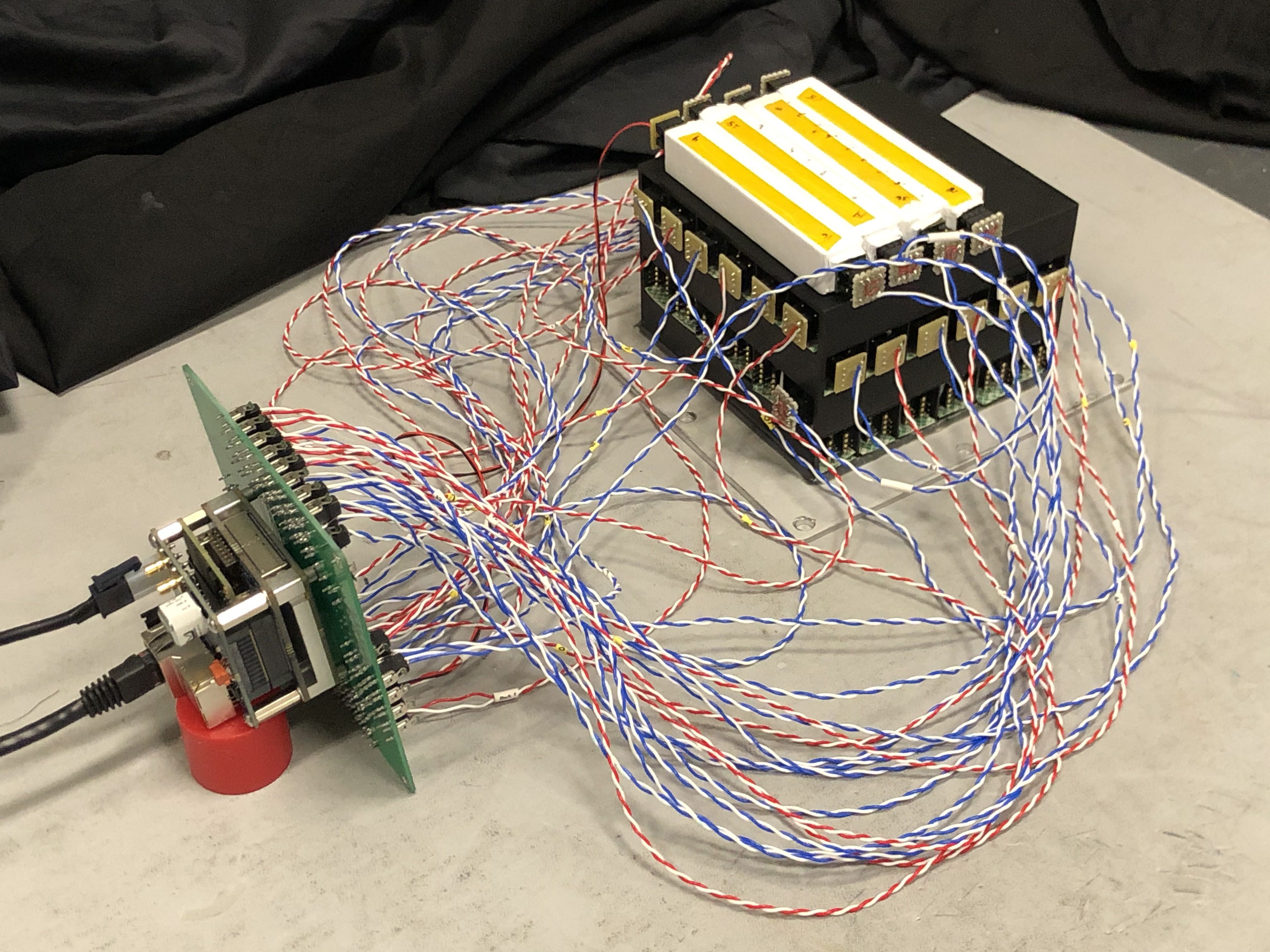}
    \caption{The AMEGO prototype High Energy Calorimeter consists of 30 1.7$\times$1.7$\times$10~cm$^{3}$ CsI bars arranged in a 5-layer hodoscopic pattern. The bars are read out on each end with a 2$\times$2 SiPM. An exploded view of one of the CsI bars removed from the full assembly is shown on the left. The photo on the right depicts 28 bars being tested with the IDEAS ROSSPAD; four wrapped crystals can be seen sitting on top of the 3D printed plastic housing.}
    \label{fig:compairCsI}
\end{figure}

The 30 CsI bars are housed in a 3D printed plastic structure, as shown in Figure~\ref{fig:compairCsI}. 
The 60 SiPM channels (2 channels per bar) are read out with the new IDEAS 64-channel ROSSPAD (Read Out System for Silicon Photomultiplier Avalanche Diodes), which contains 4 SIPHRA ASICs. 
The prototype High Energy Calorimeter also uses an Arduino Due to interface with the Trigger Module, since the ROSSPAD cannot accept the 8-bit event ID that allows the subsystems to determine coincident triggers across the different subsystems.
The prototype CsI calorimeter has been interfaced and tested with the ROSSPAD readout, and calibration of the CsI bars with collimated gamma-ray sources and atmospheric muons is underway. See Ref.~\citenum{Woolf2018} for a detailed description of the AMEGO prototype High Energy Calorimeter.

\subsection{Anti-Coincidence Detector Prototype}
\label{sec:ComPairACD}

The prototype ACD, which is being designed, tested, and built at GSFC, consists of 5 panels of 15~mm thick Eljen EJ-208 plastic scintillator that surround the four sides and the top of the instrument. 
Each panel has wavelength shifting bars across two adjacent edges with air gap coupling, and the corner where they meet is machined flat and read out by 2$\times$2 arrays of SensL C-series 6mm SiPMs.
The prototype ACD data acquisition uses the CAEN DT5550W readout system with 2 CITIROC ASICs.
The panels are wrapped to maximize light collection and then enclosed in a thin aluminium mechanical structure. 
Since the ACD is not needed for the beam test, the development of the subsystem has been delayed relative to the rest of the instrument. 
Currently, simulations and measurements with a test assembly are being performed to determine the uniformity of the wavelength shifting bar readout scheme prior to assembly of the balloon instrument.

\subsection{Trigger Module and Data Acquisition System}
\label{sec:ComPairTrigger}

Each of the AMEGO prototype subsystems is able to operate as a stand-alone detector; however, the goal of the AMEGO prototype work is to have the subsystems work together as a Compton and pair-conversion telescope. 
In the current design, each subsystem has its own data acquisition (DAQ), and the Trigger Module is used to determine when coincidence conditions between the subsystems are satisfied. 
The Trigger Module then sends out a trigger acknowledgement and an 8-bit Event ID to all subsystem DAQ systems. 
The event ID and clock information can then be used for off-line event reconstruction for the full prototype instrument.

\begin{figure}[tb]
    \centering
    \includegraphics[width=3.5in]{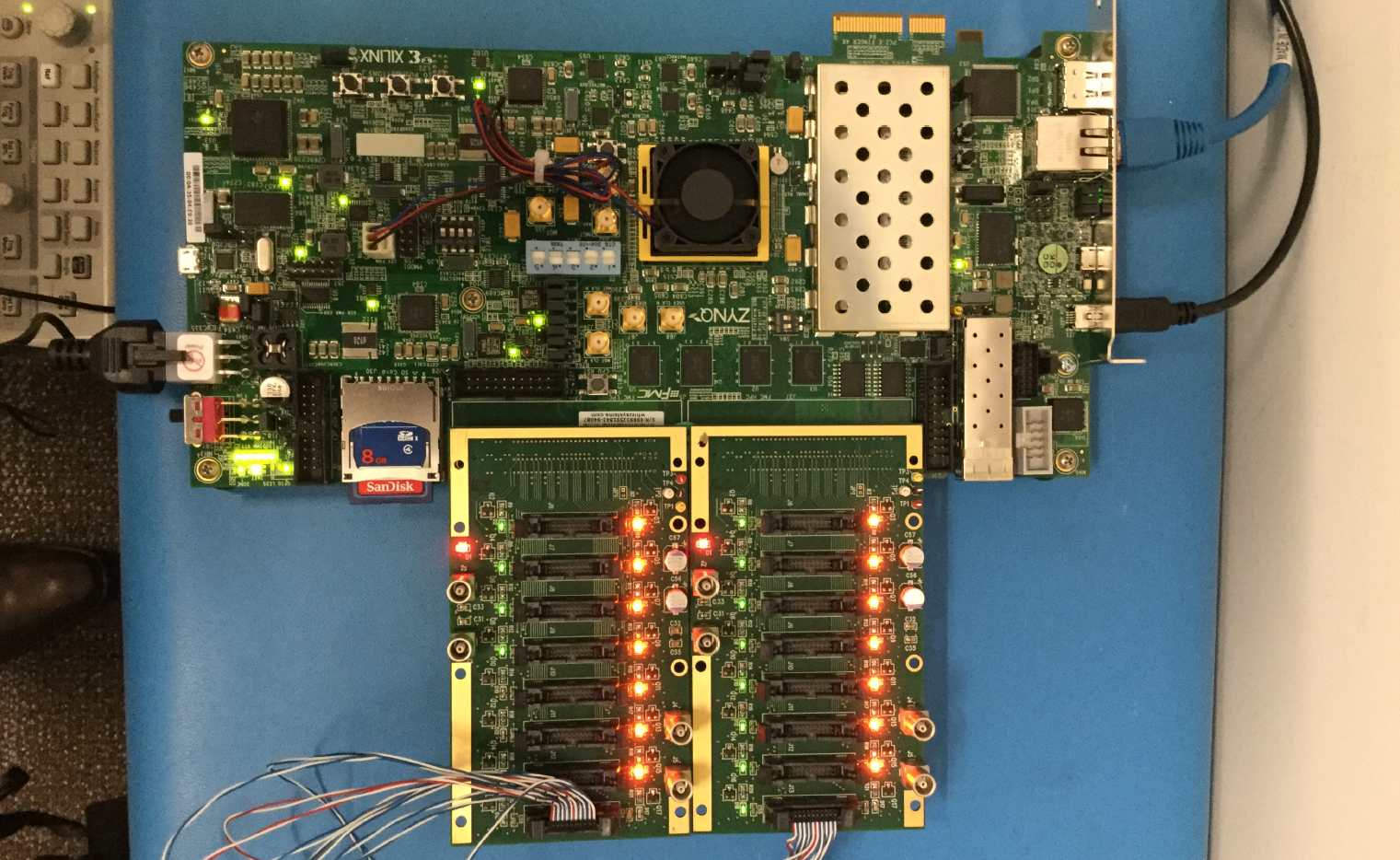}
    \caption{The AMEGO prototype Trigger Module consists of a Xilinx FPGA Evaluation Board and two custom connector boards to send and receive signals from each of the subsystem FEE boards.}
    \label{fig:ComPairTrigger}
\end{figure}

The Trigger Module, designed and built at GSFC, uses the Xilinx FPGA Evaluation Board (ZC706) with two separate custom connector boards to distribute and receive signals from each FEE board; see Fig.~\ref{fig:ComPairTrigger}. The connector boards allow for 14 separate FEE connections: the ACD and Low Energy Calorimeter each use one, the High Energy Calorimeter uses two, and the 10 layers of the Tracker take up 10 connections. 

The Trigger Module currently has 16 different trigger conditions that account for all possible coincidence outcomes in the four subsystems of the instrument. 
The ideal trigger condition corresponds to two hits in the Tracker (coincidence between X-Y (top/bottom) strips in a single detector), and one in either calorimeter; however, other trigger conditions are being explored.
The Tracker itself has 3 different coincidence models: 1) only one side of one detector has a hit (very susceptible to noise), 2) coincidence between X-Y on a single Tracker layer, and 3) any two Tracker layers have a hit on one side of a detector. 
The variety of coincidence modes, in addition to a pre-scalings option which can be applied to the different trigger modes, allow for a flexible system which can accommodate noisy and dead strips. 
The Trigger Module has been tested with mock detector inputs and has been fully integrated with the Low Energy Calorimeter. 
Tests with the High Energy Calorimeter and Tracker are expected to take place over the next few months. See Ref.~\citenum{sasakiSPIE} for a detailed description of the AMEGO prototype Trigger Module.

The software pipeline for the AMEGO prototype is under development on the instrument level
and the team is working towards the development of a single, system-wide DAQ and analysis system for the prototype balloon flight.

\subsection{Integration and Testing}

Each of the AMEGO prototype subsystems is housed in an individual aluminum box, as shown in Fig.~\ref{fig:ComPair}. 
The distances between each of the subsystems and the layers of the Tracker were minimized in an attempt to maximize the effective area and FOV of the instrument. 
The CsI High Energy Calorimeter sits 45~mm below the CZT Low Energy Calorimeter, which is 35~mm below the last Si Tracker layer. 
In mechanical integration, the edges of the subsystem boxes bolt together, as shown in Fig.~\ref{fig:ComPair}. 
The precise alignment needed between the Tracker layers and the CZT calorimeter is achieved with alignment rods which pass through two corners of the Tracker carrier boards and into the CZT Low Energy Calorimeter box. 
Alignment between the CsI High Energy Calorimeter and the Low Energy Calorimeter does not need to be as accurate due to the poor position resolution of the CsI bars.

While the subsystems are currently being developed and tested individually, they will be integrated, tested, and calibrated as a single instrument at GSFC.
This includes basic functionality tests and a full-system coincidence test with the Trigger Module to confirm timing diagrams and trigger logic. After a calibration of each subsystem is performed, an off-line analysis will be done to confirm the sucessful reconstructed Compton events and muon tracks. A full calibration of the prototype's reconstructed energy resolution, angular resolution, and effective area will follow.

The major test of the instrument will be at the free-electron laser HIGS beam facility at Duke University. The instrument will be tested at 5 different energies between 1.5 MeV to 29.5 MeV, which spans the difficult Compton and pair conversion cross-over regime. The beam has a standard energy resolution of 3-4\% and a tunable flux that is expected to be set to $\sim$1 kHz. HIGS provides a polarized beam, but we will also test with both a polarized and unpolarized beam at the lowest energies to measure the polarization sensitivity of the prototype instrument.

Unfortunately, the COVID-19 pandemic has impacted the schedule of the AMEGO prototype integration, but the team is actively working towards delivering the instrument to the HIGS beam test in 2021, with a conventional balloon flight planned from Fort Sumner, NM in 2022.

\section{CONCLUSION}

The AMEGO Probe-class mission will enable multimessenger astrophysics in the next decade. 
Using four subsystems that together measure Compton and pair events, AMEGO is sensitive across 4 orders-of-magnitude in energy.
A prototype instrument is currently being developed to validate the design of AMEGO, and tests are underway.

\acknowledgments 
 
The author would like to acknowledge the significant work from Julie McEnery, Regina Caputo, Jeremy Perkins, and Judy Racusin that went into developing the AMEGO Request for Information (RFI) response to the Astro2020 Decadal Survey that served as a basis for much of the work presented here\cite{AMEGORFI}.
Iker Liceaga Indart created all mechanical drawings for the AMEGO RFI and these proceedings.
The author would also like to thank the ComPair hardware team, namely the subsystem leads Rich Wulf, Alex Moiseev, and Jeremy S. Perkins.

\bibliography{report} 

\begin{thebibliography}{10}

\bibitem{sn1987a}
S.~M. {Matz, et al.}, ``Gamma-ray line emission from sn1987a,'' {\em
  Nature}~{\bf 331}(6155),  416--418 (1988).

\bibitem{grb170817}
A.~{Goldstein, et al.}, ``{An Ordinary Short Gamma-Ray Burst with Extraordinary
  Implications: Fermi-GBM Detection of GRB 170817A},'' {\em \apjl}~{\bf 848},
  L14 (Oct. 2017).

\bibitem{txs0506}
{IceCube Collaboration, et al.}, ``{Multimessenger observations of a flaring
  blazar coincident with high-energy neutrino IceCube-170922A},'' {\em
  Science}~{\bf 361},  eaat1378 (July 2018).

\bibitem{AMEGOBAAS}
J.~{McEnery, et al.}, ``{All-sky Medium Energy Gamma-ray Observatory: Exploring
  the Extreme Multimessenger Universe},'' in [{\em Bulletin of the American
  Astronomical Society}{\nolinebreak\hspace{0.1em}]},   {\bf 51},  245 (Sept.
  2019).

\bibitem{AMEGORFI}
J.~{McEnery, et al.}, ``{AMEGO: A Multimessenger Mission for the Extreme
  Universe - Response to Astro2020 Decadal Request for Information}.''
  \url{https://asd.gsfc.nasa.gov/amego/files/AMEGO_Decadal_RFI.pdf} (2019).

\bibitem{zoglauer2003}
A.~Zoglauer and G.~Kanbach, ``{Doppler Broadening as a Lower Limit to the
  Angular Resolution of Next Generation Compton Telescopes},'' {\em Proceedings
  of SPIE}~{\bf 4851} (2003).

\bibitem{fermi}
W.~B. {Atwood, et al.}, ``{The Large Area Telescope on the Fermi Gamma-Ray
  Space Telescope Mission},'' {\em \apj}~{\bf 697},  1071--1102 (June 2009).

\bibitem{mega}
G.~{Kanbach}, R.~{Andritschke}, F.~{Schopper}, V.~{Sch{\"o}nfelder},
  A.~{Zoglauer}, {\em et~al.}, ``{The MEGA project},'' {\em \nar}~{\bf 48},
  275--280 (Feb 2004).

\bibitem{tigre}
T.~J. {O'Neill}, A.~{Aky{\"u}z}, D.~{Bhattacharya}, M.~{Polsen}, J.~{Samimi},
  and A.~{Zych}, ``{The TIGRE gamma-ray telescope},'' in [{\em Gamma 2001:
  Gamma-Ray Astrophysics}{\nolinebreak\hspace{0.1em}]},  S.~{Ritz},
  N.~{Gehrels}, and C.~R. {Shrader}, eds., {\em American Institute of Physics
  Conference Series} {\bf 587},  882--886 (Oct. 2001).

\bibitem{eastrogam}
A.~{De Angelis, et al.}, ``{The e-ASTROGAM mission. Exploring the extreme
  Universe with gamma rays in the MeV - GeV range},'' {\em Experimental
  Astronomy}~{\bf 44},  25--82 (Oct. 2017).

\bibitem{ams02}
A.~{Kounine, et al.}, ``{The Alpha Magnetic Spectrometer on the International
  Space Station},'' {\em International Journal of Modern Physics E}~{\bf 21},
  1230005 (Aug. 2012).

\bibitem{astrohHXI}
G.~{Sato, et al.}, ``{The Hard X-ray Imager (HXI) for the ASTRO-H Mission},''
  in [{\em Space Telescopes and Instrumentation 2014: Ultraviolet to Gamma
  Ray}{\nolinebreak\hspace{0.1em}]},  T.~Takahashi, J.-W.~A. den Herder, and
  M.~Bautz, eds.,  {\bf 9144},  673 -- 683, International Society for Optics
  and Photonics, SPIE (2014).

\bibitem{pamela}
S.~{Straulino} and {Pamela Tracker Collaboration}, ``{The PAMELA silicon
  tracker},'' {\em Nuclear Instruments and Methods in Physics Research A}~{\bf
  530},  168--172 (Sept. 2004).

\bibitem{bolotnikov2020}
A.~E. {Bolotnikov, et al.}, ``{A 4 {\texttimes} 4 array module of
  position-sensitive virtual Frisch-grid CdZnTe detectors for gamma-ray imaging
  spectrometers},'' {\em Nuclear Instruments and Methods in Physics Research
  A}~{\bf 954},  161036 (Feb. 2020).

\bibitem{bolotnikov2016}
A.~E. {Bolotnikov, et al.}, ``{CdZnTe position-sensitive drift detectors with
  thicknesses up to 5 cm},'' {\em Applied Physics Letters}~{\bf 108},  093504
  (Feb. 2016).

\bibitem{bolotnikov2014spectralcorrection}
A.~E. {Bolotnikov, et al.}, ``{Use of high-granularity position sensing to
  correct response non-uniformities of CdZnTe detectors},'' {\em Applied
  Physics Letters}~{\bf 104},  263503 (June 2014).

\bibitem{bat}
S.~D. {Barthelmy, et al.}, ``{The Burst Alert Telescope (BAT) on the SWIFT
  Midex Mission},'' {\em Space Science Reviews}~{\bf 120},  143--164 (Oct.
  2005).

\bibitem{astrosat}
V.~{Bhalerao, et al.}, ``{The Cadmium Zinc Telluride Imager on AstroSat},''
  {\em Journal of Astrophysics and Astronomy}~{\bf 38},  31 (June 2017).

\bibitem{nustar}
F.~A. {Harrison, et al.}, ``{The Nuclear Spectroscopic Telescope Array
  ({NuSTAR}) High-Energy X-Ray Mission},'' {\em The Astrophysical Journal}~{\bf
  770},  103 (May 2013).

\bibitem{Woolf2018}
R.~S. {Woolf}, J.~E. {Grove}, B.~F. {Phlips}, and E.~A. {Wulf}, ``{Development
  of a CsI:Tl calorimeter subsystem for the All-Sky Medium-Energy Gamma-Ray
  Observatory (AMEGO)},'' in [{\em 2018 IEEE Nuclear Science Symposium and
  Medical Imaging Conference Proceedings
  (NSS/MIC)}{\nolinebreak\hspace{0.1em}]},   1--6 (2018).

\bibitem{siri}
L.~J. {Mitchell}, B.~F. {Phlips}, J.~E. {Grove}, T.~{Finne},
  M.~{Johnson-Rambert}, and W.~N. {Johnson}, ``{Strontium Iodide Radiation
  Instrument (SIRI) – Early On-Orbit Results},'' in [{\em 2019 IEEE Nuclear
  Science Symposium and Medical Imaging Conference
  (NSS/MIC)}{\nolinebreak\hspace{0.1em}]},   1--9 (2019).

\bibitem{burstcube}
J.~{Smith} and {BurstCube Collaboration}, ``{BurstCube: Mission Concept,
  Performance, and Status},'' in [{\em 36th International Cosmic Ray Conference
  (ICRC2019)}{\nolinebreak\hspace{0.1em}]},  {\em International Cosmic Ray
  Conference} {\bf 36},  604 (July 2019).

\bibitem{extp}
S.~N. {Zhang, et al.,}, ``{eXTP: Enhanced X-ray Timing and Polarization
  mission},'' in [{\em Space Telescopes and Instrumentation 2016: Ultraviolet
  to Gamma Ray}{\nolinebreak\hspace{0.1em}]},  J.-W.~A. den Herder,
  T.~Takahashi, and M.~Bautz, eds.,  {\bf 9905},  505 -- 520, International
  Society for Optics and Photonics, SPIE (2017).

\bibitem{calet}
Y.~{Asaoka, et al.}, ``The {CALorimetric} electron telescope ({CALET}) on the
  international space station: Results from the first two years on orbit,''
  {\em Journal of Physics: Conference Series}~{\bf 1181},  012003 (feb 2019).

\bibitem{2006NewAR..50..629Z}
A.~{Zoglauer}, R.~{Andritschke}, and F.~{Schopper}, ``{{MEGAlib The Medium
  Energy Gamma-ray Astronomy Library}},'' {\em New Astronomy Reviews}~{\bf 50},
   629--632 (Oct 2006).

\bibitem{cosi}
J.~{Tomsick, et al.}, ``{The Compton Spectrometer and Imager},'' in [{\em
  Bulletin of the American Astronomical Society}{\nolinebreak\hspace{0.1em}]},
   {\bf 51},  98 (Sept. 2019).

\bibitem{Comptel:1993}
V.~Schoenfelder {\em et~al.}, ``{Instrument description and performance of the
  Imaging Gamma-Ray Telescope COMPTEL aboard the Compton Gamma-Ray
  Observatory},'' {\em Astrophys. J. Suppl.}~{\bf 86},  657 (1993).

\bibitem{egret}
D.~J. {Thompson, et al.}, ``{Calibration of the Energetic Gamma-Ray Experiment
  Telescope (EGRET) for the Compton Gamma-Ray Observatory},'' {\em \apjs}~{\bf
  86},  629 (June 1993).

\bibitem{spi}
G.~{Vedrenne, et al.}, ``{SPI: The spectrometer aboard INTEGRAL},'' {\em
  \aap}~{\bf 411},  L63--L70 (Nov. 2003).

\bibitem{Zoglauer2020}
A.~{Zoglauer}, ``{Using Deep Learning for the Event Reconstruction of Combined
  Compton-scattering and Pair-creation Telescopes},'' in [{\em American
  Astronomical Society Meeting Abstracts \#235}{\nolinebreak\hspace{0.1em}]},
  {\em American Astronomical Society Meeting Abstracts} {\bf 235},  372.21
  (Jan. 2020).

\bibitem{griffinSPIE}
S.~{Griffin}, C.~A. {Kierans}, L.~{Parker}, A.~{Schoenwald}, P.~{Shawhan},
  R.~{Caputo}, J.~{McEnery}, J.~S. {Perkins}, and {ComPair Team}, ``{Current
  Status of the ComPair Silicon Tracker},'' {\em These Proceedings} ,
  11444--323.

\bibitem{Bolotnikov2007}
A.~E. {Bolotnikov, et al.}, ``{Optimization of virtual Frisch-grid CdZnTe
  detector designs for imaging and spectroscopy of gamma rays},'' in [{\em Hard
  X-Ray and Gamma-Ray Detector Physics IX}{\nolinebreak\hspace{0.1em}]},  R.~B.
  {James}, A.~{Burger}, and L.~A. {Franks}, eds., {\em Society of Photo-Optical
  Instrumentation Engineers (SPIE) Conference Series} {\bf 6706},  670603
  (Sept. 2007).

\bibitem{Bolotnikov2014}
A.~E. {Bolotnikov, et al.}, ``{Use of high-granularity position sensing to
  correct response non-uniformities of CdZnTe detectors},'' {\em Applied
  Physics Letters}~{\bf 104},  263503 (June 2014).

\bibitem{VERNON20191}
E.~{Vernon, et al.}, ``Front-end asic for spectroscopic readout of virtual
  frisch-grid czt bar sensors,'' {\em Nuclear Instruments and Methods in
  Physics Research Section A: Accelerators, Spectrometers, Detectors and
  Associated Equipment}~{\bf 940},  1 -- 11 (2019).

\bibitem{hays2020}
A.~{Moiseev}, A.~{Bolotnikov}, C.~{Kierans}, E.~A. {Hays}, and D.~{Thompson},
  ``{Modular Position-sensitive High-resolution Calorimeter for Use in Space
  Gamma-ray Instruments Based on Virtual Frisch-grid CdZnTe Detectors},'' in
  [{\em 36th International Cosmic Ray Conference
  (ICRC2019)}{\nolinebreak\hspace{0.1em}]},  {\em International Cosmic Ray
  Conference} {\bf 36},  584 (July 2019).

\bibitem{sasakiSPIE}
M.~{Sasaki} and {ComPair Team}, ``{Trigger system for the ComPair
  instrument},'' {\em These Proceedings} ,  11444--214.

\end{thebibliography}
\bibliographystyle{spiebib} 

\end{document}